\documentclass[journal,10pt]{IEEEtran}
\IEEEoverridecommandlockouts
\usepackage[utf8]{inputenc}
\usepackage[T1]{fontenc}
\usepackage{mathtools,gensymb,eurosym,url,soul} % provides amsmath, generic symbols, URL management and strikeout
\usepackage{subcaption}% \usepackage[font=footnotesize,textfont=footnotesize]{subcaption}
\usepackage[colorinlistoftodos,prependcaption,textsize=tiny]{todonotes}
\usepackage{longtable,booktabs,array}
\usepackage[switch]{lineno}
\usepackage{acronym}
\usepackage{graphicx}
\usepackage{xcolor}%[usenames,dvipsnames,svgnames,table]{xcolor}
\usepackage{cite}
\usepackage{xr} % for referring to the paper document

\DeclareCaptionType{equ}[][]
\usepackage{hyperref}

\hypersetup{colorlinks=true, breaklinks=true, %pagebackref=true,
  urlcolor=blue, linkcolor=blue,anchorcolor=blue,citecolor=blue,
  hypertexnames=true, final=true, 
  pdfpagemode = UseNone, %FullScreen, %UseThumbs, %UseOutlines,
  pdfauthor = {},
  pdftitle = {},   
  pdfsubject = {},
  pdfkeywords = {}
}

\graphicspath{{img/} {images/} {./}}
\newcolumntype{P}[1]{>{\centering\arraybackslash}p{#1}}
\newcolumntype{M}[1]{>{\centering\arraybackslash}m{#1}}

\def\BibTeX{{\rm B\kern-.05em{\sc i\kern-.025em b}\kern-.08em
    T\kern-.1667em\lower.7ex\hbox{E}\kern-.125emX}}

\begin{document}

\newcommand*{\unit}[1]{\ensuremath{\mathrm{\,#1}}}

\acrodef{ADC}[ADC]{Analog to Digital Converter}
\acrodef{ADEXP}[AdExp-I\&F]{Adaptive-Exponential Integrate and Fire}
\acrodef{ADM}[ADM]{Asynchronous Delta Modulator}
\acrodef{AER}[AER]{Address-Event Representation}
\acrodef{AEX}[AEX]{AER EXtension board}
\acrodef{AE}[AE]{Address-Event}
\acrodef{AFM}[AFM]{Atomic Force Microscope}
\acrodef{AGC}[AGC]{Automatic Gain Control}
\acrodef{AMDA}[AMDA]{AER Motherboard with D/A converters}
\acrodef{ANN}[ANN]{Artificial Neural Network}
\acrodef{API}[API]{Application Programming Interface}
\acrodef{APMOM}[APMOM]{Alternate Polarity Metal On Metal}
\acrodef{ARM}[ARM]{Advanced RISC Machine}
\acrodef{ASIC}[ASIC]{Application Specific Integrated Circuit}
\acrodef{AdExp}[AdExp-IF]{Adaptive Exponential Integrate-and-Fire}
\acrodef{BCM}[BMC]{Bienenstock-Cooper-Munro}
\acrodef{BD}[BD]{Bundled Data}
\acrodef{BEOL}[BEOL]{Back-end of Line}
\acrodef{BG}[BG]{Bias Generator}
\acrodef{BMI}[BMI]{Brain-Machince Interface}
\acrodef{BTB}[BTB]{band-to-band tunnelling}
\acrodef{CAD}[CAD]{Computer Aided Design}
\acrodef{CAM}[CAM]{Content Addressable Memory}
\acrodef{CAVIAR}[CAVIAR]{Convolution AER Vision Architecture for Real-Time}
\acrodef{CA}[CA]{Cortical Automaton}
\acrodef{CCN}[CCN]{Cooperative and Competitive Network}
\acrodef{CDR}[CDR]{Clock-Data Recovery}
\acrodef{CFC}[CFC]{Current to Frequency Converter}
\acrodef{CHP}[CHP]{Communicating Hardware Processes}
\acrodef{CMIM}[CMIM]{Metal-insulator-metal Capacitor}
\acrodef{CML}[CML]{Current Mode Logic}
\acrodef{CMOL}[CMOL]{Hybrid CMOS nanoelectronic circuits}
\acrodef{CMOS}[CMOS]{Complementary Metal-Oxide-Semiconductor}
\acrodef{CNN}[CCN]{Convolutional Neural Network}
\acrodef{COTS}[COTS]{Commercial Off-The-Shelf}
\acrodef{CPG}[CPG]{Central Pattern Generator}
\acrodef{CPLD}[CPLD]{Complex Programmable Logic Device}
\acrodef{CPU}[CPU]{Central Processing Unit}
\acrodef{CSM}[CSM]{Cortical State Machine}
\acrodef{CSP}[CSP]{Constraint Satisfaction Problem}
\acrodef{CTXCTL}[CTXCTL]{CortexControl}
\acrodef{CV}[CV]{Coefficient of Variation}
\acrodef{DAC}[DAC]{Digital to Analog Converter}
\acrodef{DAS}[DAS]{Dynamic Auditory Sensor}
\acrodef{DAVIS}[DAVIS]{Dynamic and Active Pixel Vision Sensor}
\acrodef{DBN}[DBN]{Deep Belief Network}
\acrodef{DFA}[DFA]{Deterministic Finite Automaton}
\acrodef{DIBL}[DIBL]{drain-induced-barrier-lowering}
\acrodef{DI}[DI]{delay insensitive}
\acrodef{DMA}[DMA]{Direct Memory Access}
\acrodef{DNF}[DNF]{Dynamic Neural Field}
\acrodef{DNN}[DNN]{Deep Neural Network}
\acrodef{DOF}[DOF]{Degrees of Freedom}
\acrodef{DPE}[DPE]{Dynamic Parameter Estimation}
\acrodef{DPI}[DPI]{Differential Pair Integrator}
\acrodef{DRAM}[DRAM]{Dynamic Random Access Memory}
\acrodef{DRRZ}[DR-RZ]{Dual-Rail Return-to-Zero}
\acrodef{DR}[DR]{Dual Rail}
\acrodef{DSP}[DSP]{Digital Signal Processor}
\acrodef{DVS}[DVS]{Dynamic Vision Sensor}
\acrodef{DYNAP}[DYNAP]{Dynamic Neuromorphic Asynchronous Processor}
\acrodef{EBL}[EBL]{Electron Beam Lithography}
\acrodef{EDVAC}[EDVAC]{Electronic Discrete Variable Automatic Computer}
\acrodef{EEG}[EEG]{electroencephalography}
\acrodef{EIN}[EIN]{Excitatory-Inhibitory Network}
\acrodef{EM}[EM]{Expectation Maximization}
\acrodef{EPSC}[EPSC]{Excitatory Post-Synaptic Current}
\acrodef{EPSP}[EPSP]{Excitatory Post-Synaptic Potential}
\acrodef{EZ}[EZ]{Epileptogenic Zone}
\acrodef{FDSOI}[FDSOI]{Fully-Depleted Silicon on Insulator}
\acrodef{FET}[FET]{Field-Effect Transistor}
\acrodef{FFT}[FFT]{Fast Fourier Transform}
\acrodef{FI}[F-I]{Frequency-Current}
\acrodef{FPGA}[FPGA]{Field Programmable Gate Array}
\acrodef{FR}[FR]{Fast Ripple}
\acrodef{FSA}[FSA]{Finite State Automaton}
\acrodef{FSM}[FSM]{Finite State Machine}
\acrodef{GIDL}[GIDL]{gate-induced-drain-leakage}
\acrodef{GOPS}[GOPS]{Giga-Operations per Second}
\acrodef{GPU}[GPU]{Graphical Processing Unit}
\acrodef{GUI}[GUI]{Graphical User Interface}
\acrodef{HAL}[HAL]{Hardware Abstraction Layer}
\acrodef{HFO}[HFO]{High Frequency Oscillation}
\acrodef{HH}[H\&H]{Hodgkin \& Huxley}
\acrodef{HMM}[HMM]{Hidden Markov Model}
\acrodef{HRS}[HRS]{High-Resistive State}
\acrodef{HR}[HR]{Human Readable}
\acrodef{HSE}[HSE]{Handshaking Expansion}
\acrodef{HW}[HW]{Hardware}
\acrodef{ICT}[ICT]{Information and Communication Technology}
\acrodef{IC}[IC]{Integrated Circuit}
\acrodef{IEEG}[iEEG]{intracranial electroencephalography}
\acrodef{IF2DWTA}[IF2DWTA]{Integrate \& Fire 2--Dimensional WTA}
\acrodef{IFSLWTA}[IFSLWTA]{Integrate \& Fire Stop Learning WTA}
\acrodef{IF}[I\&F]{Integrate-and-Fire}
\acrodef{IMU}[IMU]{Inertial Measurement Unit}
\acrodef{INCF}[INCF]{International Neuroinformatics Coordinating Facility}
\acrodef{INI}[INI]{Institute of Neuroinformatics}
\acrodef{IO}[I/O]{Input/Output}
\acrodef{IPSC}[IPSC]{Inhibitory Post-Synaptic Current}
\acrodef{IPSP}[IPSP]{Inhibitory Post-Synaptic Potential}
\acrodef{IP}[IP]{Intellectual Property}
\acrodef{ISI}[ISI]{Inter-Spike Interval}
\acrodef{IoT}[IoT]{Internet of Things}
\acrodef{JFLAP}[JFLAP]{Java - Formal Languages and Automata Package}
\acrodef{LEDR}[LEDR]{Level-Encoded Dual-Rail}
\acrodef{LFP}[LFP]{Local Field Potential}
\acrodef{LLC}[LLC]{Low Leakage Cell}
\acrodef{LNA}[LNA]{Low-Noise Amplifier}
\acrodef{LPF}[LPF]{Low Pass Filter}
\acrodef{LRS}[LRS]{Low-Resistive State}
\acrodef{LSM}[LSM]{Liquid State Machine}
\acrodef{LTD}[LTD]{Long Term Depression}
\acrodef{LTI}[LTI]{Linear Time-Invariant}
\acrodef{LTP}[LTP]{Long Term Potentiation}
\acrodef{LTU}[LTU]{Linear Threshold Unit}
\acrodef{LUT}[LUT]{Look-Up Table}
\acrodef{LVDS}[LVDS]{Low Voltage Differential Signaling}
\acrodef{MCMC}[MCMC]{Markov-Chain Monte Carlo}
\acrodef{MEMS}[MEMS]{Micro Electro Mechanical System}
\acrodef{MFR}[MFR]{Mean Firing Rate}
\acrodef{MIM}[MIM]{Metal Insulator Metal}
\acrodef{MLP}[MLP]{Multilayer Perceptron}
\acrodef{MOSCAP}[MOSCAP]{Metal Oxide Semiconductor Capacitor}
\acrodef{MOSFET}[MOSFET]{Metal Oxide Semiconductor Field-Effect Transistor}
\acrodef{MOS}[MOS]{Metal Oxide Semiconductor}
\acrodef{MRI}[MRI]{Magnetic Resonance Imaging}
\acrodef{NDFSM}[NDFSM]{Non-deterministic Finite State Machine} 
\acrodef{ND}[ND]{Noise-Driven}
\acrodef{NEF}[NEF]{Neural Engineering Framework}
\acrodef{NHML}[NHML]{Neuromorphic Hardware Mark-up Language}
\acrodef{NIL}[NIL]{Nano-Imprint Lithography}
\acrodef{NMDA}[NMDA]{N-Methyl-D-Aspartate}
\acrodef{NME}[NE]{Neuromorphic Engineering}
\acrodef{NN}[NN]{Neural Network}
\acrodef{NRZ}[NRZ]{Non-Return-to-Zero}
\acrodef{NSM}[NSM]{Neural State Machine}
\acrodef{OR}[OR]{Operating Room}
\acrodef{OTA}[OTA]{Operational Transconductance Amplifier}
\acrodef{PCB}[PCB]{Printed Circuit Board}
\acrodef{PCHB}[PCHB]{Pre-Charge Half-Buffer}
\acrodef{PCM}[PCM]{Phase Change Memory}
\acrodef{PE}[PE]{Phase Encoding}
\acrodef{PFA}[PFA]{Probabilistic Finite Automaton}
\acrodef{PFC}[PFC]{prefrontal cortex}
\acrodef{PFM}[PFM]{Pulse Frequency Modulation}
\acrodef{PR}[PR]{Production Rule}
\acrodef{PSC}[PSC]{Post-Synaptic Current}
\acrodef{PSP}[PSP]{Post-Synaptic Potential}
\acrodef{PSTH}[PSTH]{Peri-Stimulus Time Histogram}
\acrodef{QDI}[QDI]{Quasi Delay Insensitive}
\acrodef{RAM}[RAM]{Random Access Memory}
\acrodef{RA}[RA]{Resected Area}
\acrodef{RDF}[RDF]{random dopant fluctuation}
\acrodef{RELU}[ReLu]{Rectified Linear Unit}
\acrodef{RLS}[RLS]{Recursive Least-Squares}
\acrodef{RMSE}[RMSE]{Root Mean Squared-Error}
\acrodef{RMS}[RMS]{Root Mean Squared}
\acrodef{RNN}[RNN]{Recurrent Neural Networks}
\acrodef{ROLLS}[ROLLS]{Reconfigurable On-Line Learning Spiking}
\acrodef{RRAM}[R-RAM]{Resistive Random Access Memory}
\acrodef{R}[R]{Ripples}
\acrodef{SAC}[SAC]{Selective Attention Chip}
\acrodef{SAT}[SAT]{Boolean Satisfiability Problem}
\acrodef{SCX}[SCX]{Silicon CorteX}
\acrodef{SD}[SD]{Signal-Driven}
\acrodef{SEM}[SEM]{Spike-based Expectation Maximization}
\acrodef{SLAM}[SLAM]{Simultaneous Localization and Mapping}
\acrodef{SNN}[SNN]{Spiking Neural Network}
\acrodef{SNR}[SNR]{Signal to Noise Ratio}
\acrodef{SOC}[SOC]{System-On-Chip}
\acrodef{SOI}[SOI]{Silicon on Insulator}
\acrodef{SOZ}[SOZ]{Seizure Onset Zone}
\acrodef{SP}[SP]{Separation Property}
\acrodef{SRAM}[SRAM]{Static Random Access Memory}
\acrodef{STDP}[STDP]{Spike-Timing Dependent Plasticity}
\acrodef{STD}[STD]{Short-Term Depression}
\acrodef{STP}[STP]{Short-Term Plasticity}
\acrodef{STT-MRAM}[STT-MRAM]{Spin-Transfer Torque Magnetic Random Access Memory}
\acrodef{STT}[STT]{Spin-Transfer Torque}
\acrodef{SW}[SW]{Software}
\acrodef{TCAM}[TCAM]{Ternary Content-Addressable Memory}
\acrodef{TFT}[TFT]{Thin Film Transistor}
\acrodef{TLE}[TLE]{Temporal Lobe Epilepsy}
\acrodef{USB}[USB]{Universal Serial Bus}
\acrodef{VHDL}[VHDL]{VHSIC Hardware Description Language}
\acrodef{VLSI}[VLSI]{Very Large Scale Integration}
\acrodef{VOR}[VOR]{Vestibulo-Ocular Reflex}
\acrodef{WCST}[WCST]{Wisconsin Card Sorting Test}
\acrodef{WTA}[WTA]{Winner-Take-All}
\acrodef{XML}[XML]{eXtensible Mark-up Language}
\acrodef{divmod3}[DIVMOD3]{divisibility of a number by three}
\acrodef{hWTA}[hWTA]{hard Winner-Take-All}
\acrodef{sWTA}[sWTA]{soft Winner-Take-All}

\title{Ultra-Low-Power FDSOI Neural Circuits for Extreme-Edge Neuromorphic Intelligence}

\author{
    \IEEEauthorblockN{Arianna Rubino, Can Livanelioglu, Ning Qiao, Melika Payvand~\IEEEmembership{Member,~IEEE}}, and~Giacomo Indiveri~\IEEEmembership{Senior~Member,~IEEE}
    \thanks{This paper is supported in part by the European Union's Horizon 2020 ERC project NeuroAgents (Grant No. 724295), by European Union’s Horizon 2020 research and innovation programme under grant agreement No 871737 (project BeFerroSynaptic), and by Toshiba Corporation. This research work was also partially supported by H2020 MeM-Scales project (871371) and by the ECSEL Joint Undertaking (JU) under grant agreement No 826655. The JU receives support from the European Union’s Horizon 2020 research and innovation programme and Belgium, France, Germany, Netherlands, Switzerland.}%
    \thanks{A. Rubino, N. Qiao, M. Payvand, and G. Indiveri are with the University of Zurich and ETH Zurich, Institute of Neuroinformatics, Zurich, Switzerland.
    (\mbox{e-mail}: rubinoa@ini.uzh.ch, qiaoning@ini.uzh.ch, melika@ini.uzh.ch, giacomo@ini.uzh.ch).
    C. Livanelioglu is with ETH Zurich, Zurich, Switzerland. (\mbox{e-mail}: clivaneli@student.ethz.ch).}
}

\maketitle

\begin{abstract}
  Recent years have seen an increasing interest in the development of artificial intelligence circuits and systems for edge computing applications.
  In-memory computing mixed-signal neuromorphic architectures provide promising ultra-low-power solutions for edge-computing sensory-processing applications, thanks to their ability to emulate spiking neural networks in real-time.
  The fine-grain parallelism offered by this approach allows such neural circuits to process the sensory data efficiently by adapting their dynamics to the ones of the sensed signals, without having to resort to the time-multiplexed computing paradigm of von Neumann architectures.
  To reduce power consumption even further, we present a set of mixed-signal analog/digital circuits that exploit the features of advanced \ac{FDSOI} integration processes.
  Specifically, we explore the options of advanced \ac{FDSOI} technologies to address analog design issues and optimize the design of the synapse integrator and of the adaptive neuron circuits accordingly.
  We present circuit simulation results and demonstrate the circuit's ability to produce biologically plausible neural dynamics with compact designs, optimized for the realization of large-scale spiking neural networks in neuromorphic processors.
\end{abstract}


\acresetall

\section{Introduction}

A technological revolution is in the making where more and more \ac{IoT} and edge-computing devices are being produced to sense and process signals, for example in environmental or health monitoring applications, and extract relevant information locally, without resorting to cloud computing or transferring large amounts of data to remote data centers.
This poses a serious challenge in terms of memory and power consumption requirements for \ac{IoT} systems, especially when they are expected to operate autonomously in a compact package directly on the sensed signals (i.e.,  in ``extreme-edge'' computing application scenarios).
Due to the limitations of Dennard scaling law~\cite{Davari_etal95} and the von Neumann memory bottleneck problems~\cite{Backus78,Indiveri_Liu15}, a disruptive change is needed in the development of new memory and computing technologies to be able to sustain these processing requirements under tight power and size constraints.

A promising computational paradigm that can support the ultra-low-power implementation of ``extreme-edge'' computing processing tasks is that of \acp{SNN} and attractor dynamics~\cite{Maass_Markram04,Zambrano_Bohte16,Neftci18}.
In particular, it has been shown that recurrent \acp{SNN} provide a valuable algorithmic basis for efficient processing of temporal signals: the rich dynamics of these networks are instrumental in minimizing the amount of memory resources required to process, recognize, and classify long temporal sequences of data~\cite{Lukosevicius_Jaeger09}.
In addition, recent studies suggest that longer time constants in the \ac{SNN} synapse and neuron models are very beneficial in lengthening the so-called ``fading memory'' of the recurrent network~\cite{Bellec_etal18}.
The best-suited approach to implement such networks in hardware which minimizes power consumption and area is that of using mixed-signal neuromorphic circuits~\cite{Mead90,Chicca_etal14}.
By exploiting the temporal properties of such circuits to adapt them to the temporal dynamics of the signals being processed, it is possible to implement an optimal ``matched filter'' approach that minimizes power consumption and maximizes the \ac{SNR}~\cite{Indiveri_Sandamirskaya19}.
This approach forgoes the need for storing the data and the state of the processing elements, since they operate in real-time directly on the signals being acquired by the sensor.
By combining the adaptive analog signal processing strategies of these neuromorphic circuits with digital event-based asynchronous communication schemes, it is possible to build large-scale multi-core neuromorphic processors that combine the best of both (analog and digital) world for low-power signal processing, computation, and communication.
These processors typically operate with sub-mWatt power-consumption figures and support the emulation of a wide range of \ac{SNN} models, for solving artificial intelligence tasks directly on the sensory signals, as they are acquired~(e.g., see \cite{Moradi_etal18,Bauer_etal19}).
We refer to the combination of this technology with this approach as ``\emph{extreme-edge neuromorphic intelligence}''.

The key enabling features of  neuromorphic intelligence circuits are twofold: (i) They operate in continuous real-time on sensory signals, dissipating power only when the data becomes available and (ii) their operation speed is adapted (typically slowed down) to the time-scale of the signals being processed.
For natural signals such as speech, human gestures, bio-signals, and a wide range of environmental signals, these time constants are the biologically plausible ones that range from fractions of milli-seconds to seconds.
To optimize edge-computing applications that operate on these types of input signals, a crucial precondition is to be able to support the processing of data on these different biological time scales.
It is challenging to realize a similar range of timescales using conventional technologies, as they require very large capacitors and/or big digital memory storage blocks which limits the scalability of these circuits.
As the technology nodes move toward deep sub-micron processes, the increased leakage current limits the time constants and also poses a challenge in terms of the static power consumption.
Moreover, as the technology node scales down and the transistor's channel length decreases, its parameter variations (e.g. the threshold voltage) increase, and device mismatch increases even further.

In this paper, we present sub-threshold neuron and synapse circuits that have been designed to implement large-scale multi-neuron multi-core neuromorphic computing architectures using a 22\,nm \ac{FDSOI} process.
In Section~\ref{sec:methods} we show how it is possible to implement bio-physically complex neural and synaptic dynamics using ultra-low-power analog circuits in advanced scaled processes, by analyzing the features of the 22\,nm \ac{FDSOI} technology and addressing the analog design issues that arise from the advanced scaling.
In particular, we exploit the properties of the \ac{FDSOI} technology to design a synapse circuit that can reach 6\,sec-long time constant, operating reliably with femto-Ampere currents. We also optimize the design of the \ac{FDSOI} silicon neuron circuit recently proposed in~\cite{Rubino_etal19} to further reduce the device mismatch effects.

In Section~\ref{sec:results} we present circuit simulation results highlighting how the synapse and neuron dynamics change as a function of their bias parameters. We also characterize the power consumption figures as a function of the neuron average firing rate, and quantify the effect of device mismatch with Monte Carlo analysis simulations.
\section{Methods}
\label{sec:methods}

\subsection{\ac{FDSOI} process advantages}
\label{sec:fdsoi}

Neuromorphic analog circuits typically use transistors operated in the sub-threshold or weak-inversion regime~\cite{Mead89,Vittoz96,Liu_etal02a}, using currents that range from fractions of pico-Amperes to hundreds of nano-Amperes.
Indeed, to emulate biologically plausible dynamics, with time constants of the order of tens of milli-seconds, while using small capacitors (e.g., of the order of pico-Farads), it is necessary to limit the currents to pico-Ampere amplitudes.
Furthermore, to implement circuits with even longer time-scales, for example to emulate homeostatic plasticity phenomena that last multiple seconds or tens of seconds, it would be necessary to reduce the currents even below femto-Ampere amplitudes.
However, the non-ideal transistor effects of advanced \ac{CMOS} technology nodes (e.g., \ac{DIBL}, \ac{BTB}, \ac{GIDL}, or \ac{RDF}) produce leakage currents that are well above the single pico-Ampere digits and severely limit the functionality of subthreshold neuromorphic circuits.

To design neuromorphic circuits in advanced technology nodes, one option is to resort to above-threshold circuits, either accelerating time constants by a factor of $\times$1000~\cite{Schemmel_etal17} giving up the ability to optimally process slowly changing sensory signals, or by using switched-capacitor techniques~\cite{Mayr_etal16,Folowosele_etal09,Folowosele_etal09a} increasing the complexity of the circuit design techniques and power-consumption figures; the other option is to use \ac{FDSOI} technology and exploit its features.
The \ac{FDSOI} technology introduces a variety of enhancements over the bulk \ac{CMOS} technology.
For instance, this more advanced technology utilizes a thin insulating layer of buried oxide, together with shallow trench isolation, which improves the electrical isolation between the channel and the substrate.
As a result, bulk effects such as source/drain leakage, latch-up, parasitic source/drain junction capacitances and substrate noise coupling are reduced to a significant extent.
The lower doping features lead to less threshold voltage  variation and device mismatch across the chip.
Moreover, the control over the channel and lower junction capacitances are enhanced by the addition of the second gate terminal, referred to as ``body'' or ``back-gate'', which can be reverse or forward biased to set the transistor in low leakage or high performance modes respectively.
Here we will focus on a 22\,nm \ac{FDSOI} process which  offers two main transistor types with thin and thick gate oxides. This provides the designer with a variety of transistor models that can have different threshold voltages and hence leakage currents.
In analog circuit designs it is often necessary to use non-minimum size devices (large transistors). For these cases, it will be necessary to resort to the use of thick gate oxide devices, to avoid gate-leakage due to tunneling effects.
These thick gate oxide devices can come in two flavors: Low-Threshold-Voltage LVT and Super-Low-Threshold-Voltage SLVT devices.
SLVT are typically flip-well devices featuring the same doping type for their well and channel, and the $V_{th}$ can be further reduced with \emph{forward} back-gate biasing.
LVT are conventional-well devices which have opposite doping types between channel and well and the $V_{th}$ can be further increased with \emph{reverse} back-gate biasing.
%To achieve ultra-low leakage current levels, we will use the conventional type (LVT) in combination with reverse back-gate biasing when needed.

\subsection{Key sub-circuits in 22 nm FD-SOI processes}
\label{sec:subcircuits}

In the neuron and synapse designs presented in this section, we use the 1.2 V I/O Low Threshold Transistors (LVT) at a $V_{dd}$ of 0.8\,V.
These transistors are thick gate oxide, conventional well devices with high enough $V_{th}$ which offer ultra-low leakage baseline, essential for achieving ultra-low leakage current levels.
Moreover, due to their large utilizable range of channel lengths and minimal channel doping, these transistors offer less device mismatch.

In both designs the capacitances are implemented using \ac{APMOM} structures. %Alternate Polarity Metal On Metal
The density of these devices depends on the value and on the number of used metal layers: larger capacitances can be implemented with more metal layers to attain a higher capacitance density. % of $4.5 \ - \ 5.5 fF/\mu m^2$.
However, as these devices exhibit parasitic resistance effects that scale with their area, there are severe limitations to the maximum capacitance that can be attained.

\subsubsection{The neuromorphic synapse circuit}
\label{sec:synapse}

\begin{figure*}
 \begin{center}
  \includegraphics[width= 140mm]{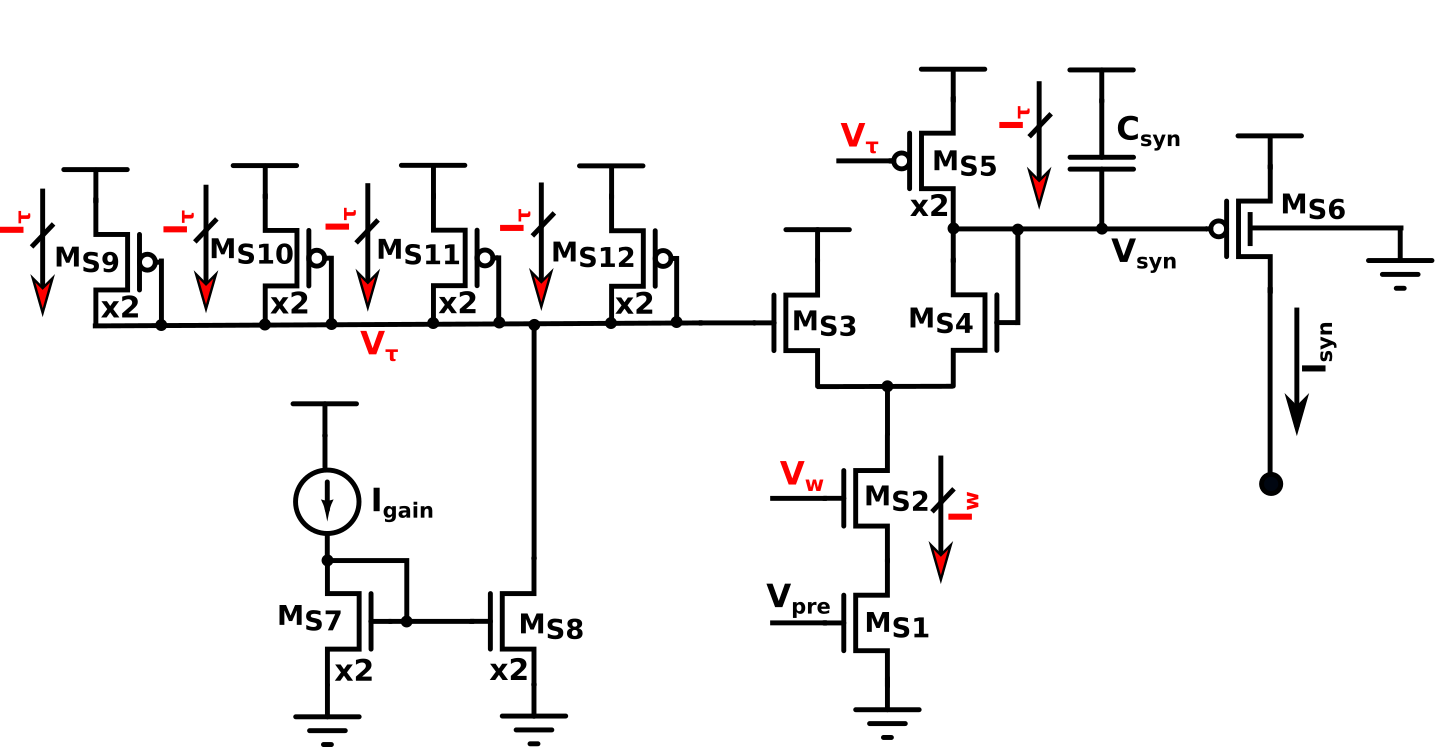}
      \caption{Schematic of the \ac{DPI} synapse circuit. Input spikes are applied to the $V_{pre}$ node.  The output \ac{EPSC} $I_{syn}$ has an amplitude proportional to $V_w$ and decays exponentially with a time constant $\tau_{syn}$ directly proportional to $C_{syn}$ and inversely proportional to $I_{\tau}$.}
  \label{fig:Syn22nm}
   \end{center}
\end{figure*}

Here we present an \ac{FDSOI} synapse circuit based on the \ac{DPI}~\cite{Bartolozzi_etal06,Bartolozzi_Indiveri07a} which models the synaptic response behavior as a first order linear system.
The schematic diagram of the circuit is shown in Fig.~\ref{fig:Syn22nm}. Input spikes applied to the $V_{pre}$ node get integrated into an \ac{EPSC} which obeys the following dynamics:
\begin{equation}
  \tau_{syn}\frac{dI_{syn}}{dt}+ I_{syn} = \frac{I_{gain}}{I_{\tau}}I_w
  \label{eq:dpi_syn}
\end{equation}

where $I_{syn}$ is the synapse output current, $I_{gain}$ is a reference current, $I_{\tau}$ is a current equivalent to $I_{gain}/4$, and $I_{w}$ is a subthreshold current set by $V_w$.
The synapse time constant $\tau_{syn}$ is defined as:
\[ \tau_{syn}\coloneqq \frac{C_{syn} \ U_T}{\kappa \ I_{\tau}} \]
where $C_{syn}$ is the \ac{DPI} capacitor, $U_T$ is the thermal voltage $KT/q$, and $\kappa$ is the subthreshold slope factor~\cite{Liu_etal02a}.
% In the design proposed here, we set the gain term to a constant ratio $I_{gain}/I_{\tau}=4$, so changing the synapse time constant parameter does not affect the maximum synapse amplitude.
To increase the circuit time constant it is therefore necessary to either increase capacitor sizes or to reduce $I_{\tau}$ currents.
Increasing capacitor sizes however is problematic, because of large area requirements and area-dependent leakage drawbacks (e.g., with APMOM structures).
So the most viable solution is to minimize the $I_{\tau}$ current.
By exploiting the features of \ac{FDSOI} technology and analog circuit design techniques, the synapse circuit presented on Fig.~\ref{fig:Syn22nm} can reliably produce $I_{\tau}$ currents of the order of femto-Amperes.
This allows the circuit to reach $\tau_{syn}$ values of up to 6\,s with compact synaptic $C_{syn}$ capacitors that have capacitance values below 1\,pF (specifically, 821\,fF with an APMOM structure of 12$\times$12\,$\mu\text{m}^2$ in our design).
Furthermore, by setting the \ac{DPI} gain term to a constant ratio ($I_{gain}/I_{\tau}=4$) we ensure that changes in the synapse time constant do not affect the maximum synapse current amplitude.

The transistors in the synapse circuit are operated with a power-supply voltage $V_{dd}=$ 0.8\,V, which is set below the nominal supply voltage of 1.2\,V, to reduce channel leakage and  power consumption.
To achieve the aimed long time constants, we configured the conventional well devices with full reverse back-gate biases ($\pm$ 2\,V).
In the schematic of Fig.~\ref{fig:Syn22nm}, all transistors are fully reverse body biased for minimum leakage operation, except for the current injecting PFET ($M_{S6}$) whose body contact was set to $gnd$ for achieving higher synaptic efficacy.
To improve the synaptic efficacy further, the NFETs of the differential pair ($M_{S3}$ and $M_{S4}$) are designed with high $\frac{W}{L}$ ratio so that they accommodate a lower $V_{gs}$ drop and provide enough $V_{ds}$ headroom for $M_{S1}$ and $M_{S2}$ to remain in saturation while $V_{syn}$ discharges to lower voltages.

To achieve even lower leakage currents and higher output impedance for the $I_{\tau}$ and $I_{gain}$ current mirrors, and allow them to operate correctly with sub-pico-Ampere currents, we adopted the self-cascoding technique proposed in~\cite{Saxena_Baker08}.
The transistor self-cascoded configuration is denoted in the figure by ``x2'' symbol.

The operation of the \ac{DPI} synapse circuit is controlled by the input $V_{pre}$, which is a pulse signal representing the incoming spikes from the previous synaptic neurons. When $V_{pre}$ is at $gnd$ there is no current flowing in the bottom branch ($M_{S1}$, $M_{S2}$ and $M_{S4}$) and $I_{\tau}$ keeps $V_{syn}$ charged at $V_{dd}$, switching OFF $M_{S6}$. When $V_{pre}$ is at $V_{dd}$, $V_{syn}$ discharges with a speed set by $I_{w}$ - $I_{\tau}$ and as $M_{S6}$ starts switching ON, the \ac{DPI} synapse circuit injects $I_{syn}$. When $V_{pre}$ returns to $gnd$, $V_{syn}$ charges back to $V_{dd}$ with a rate set by $I_{\tau}$, switching $M_{S6}$ back to the OFF state.

\subsubsection{The silicon neuron circuit}
\label{sec:neuron}

It has been recently argued that \acp{SNN} can accomplish remarkable learning and inference performance figures, if they are endowed with complex dynamics which comprise multiple and diverse time-scales~\cite{Bellec_etal18}.
To support such networks, we propose the use of the \ac{ADEXP} silicon neuron circuit~\cite{Indiveri03a,Livi_Indiveri09,Indiveri_etal11,Chicca_etal14,Rubino_etal19}.
The \ac{ADEXP} neuron model has been shown to be able to reproduce a wide range of spiking behaviors and explain a wide set of experimental measurements from pyramidal neurons \cite{Brette_Gerstner05,Naud_etal08}.
Similar to the Izhikevich neuron model~\cite{Izhikevich04}, it is a two-variable model with a ``fast'' variable that describes the dynamics of the membrane potential and includes an activation term with an exponential voltage dependence, and a ``slow'' variable that describes the spike-frequency adaptation mechanism.
This is a negative-feedback mechanism which decreases the effect of the input current to the neuron with every output spike, therefore acting as a high-pass filter which reduces the neuron firing rate in response to instantaneous increases in the input.

The equations that describe the original computational model\cite{Brette_Gerstner05} are the following:
\begin{align}
  & C\frac{dV}{dt} = -g_{L} \left(V-E_L\right) + g_L\Delta_T\cdot e^{\frac{V-V_T}{\Delta_T}} -w + I \label{eq:vmem}\\
  & \tau_w\frac{dw}{dt} = a \left(V-E_L\right) - w \label{eq:ahp}
\end{align}
where V represents the neuron membrane potential, C its membrane capacitance, $g_{L}$ the leak conductance, $E_{L}$ the resting potential,  $I$ the neuron's input current, and $w$ is the slow variable that represents the after-hyperpolarizing current of biological neurons responsible for their spike-frequency adaptation behavior~\cite{Ha_Cheong17}.
The term $\Delta_{T}$ represents the exponential slope factor,  $V_{T}$ the neuron's spiking threshold potential, $a$ the adaptation weight, and $\tau_{w}$ the adaptation time constant.
At every spike, the neuron is reset to the resting potential and the adaptation variable is increased by $a$.

\begin{figure*}
  \begin{center}
    \includegraphics[width=180mm]{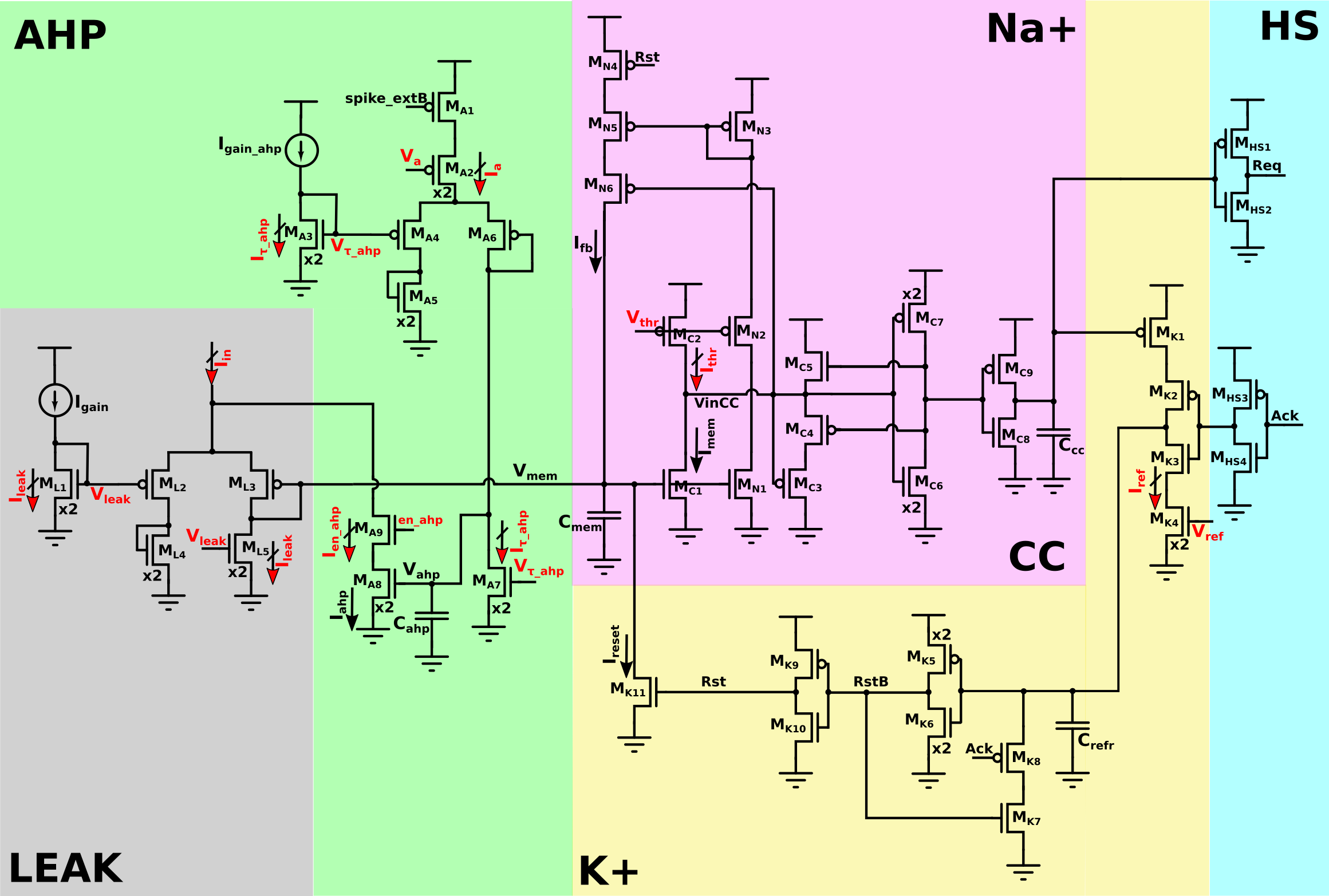}
    \caption{\ac{ADEXP} neuron circuit schematic: In grey the input \ac{DPI} \ac{LPF}, in pink the positive-feedback and the current comparator, in yellow the reset block, in light blue the handshake block and in green the spike-frequency adaptation block. Adapted from \cite{Rubino_etal19}.}
    \label{fig:AdExpIF22nm}
  \end{center}
\end{figure*}

The \ac{ADEXP} \ac{FDSOI} neuron we propose is depicted in Fig.~\ref{fig:AdExpIF22nm}.
It is a current-mode circuit, in which the currents represent the state variables.
Therefore the $V$ and $w$ variables of the computational model described by Eqs.\eqref{eq:vmem} and~\eqref{eq:ahp} are represented in the circuit by the currents $I_{mem}$ and $I_{ahp}$ respectively.
By adopting the same translinear-circuit analysis techniques used for the \ac{DPI} synapse circuit, and described in~\cite{Chicca_etal14}, and using very low leakage currents such that  $I_{leak}{\ll}I_{in}$, it is possible to express the circuit dynamics as:
\begin{align}
  C_{mem}\frac{d}{dt}I_{mem} =  & -g_LI_{mem} + g_L f(I_{mem}) \nonumber\\  % \left(1+\frac{I_{ahp}}{I_{leak}}\right)
                              & -g_LI_{ahp}   + g_LI_{in}  \label{eq:Imem}\\
  \tau_{ahp}\frac{d}{dt}I_{ahp} = & I_a - I_{ahp}   \label{eq:Iahp}
\end{align}
%\text{where} \ g_L \ \text{and} \ \tau_{ahp} \ \text{are defined as:} & \nonumber \\
where $g_L$ and $\tau_{ahp}$ are defined as:
\begin{align}
g_L\coloneqq & \frac{\kappa\,I_{leak}}{U_T} \nonumber \\
  \tau_{ahp}\coloneqq & \frac{C_{ahp}\,U_T}{\kappa\,I_{\tau_{ahp}}}, \nonumber
\end{align}
and where $I_a$ represents the adaptation weight for the slow variable, and $f(I_{mem})$ is a current produced by the positive-feedback block transistors ($M_{N1}$--$M_{N6}$) which has been shown to well fit a positive-exponent exponential function~\cite{Indiveri_etal10}.

\begin{figure}
  \begin{center}
    \includegraphics[width=0.45\textwidth]{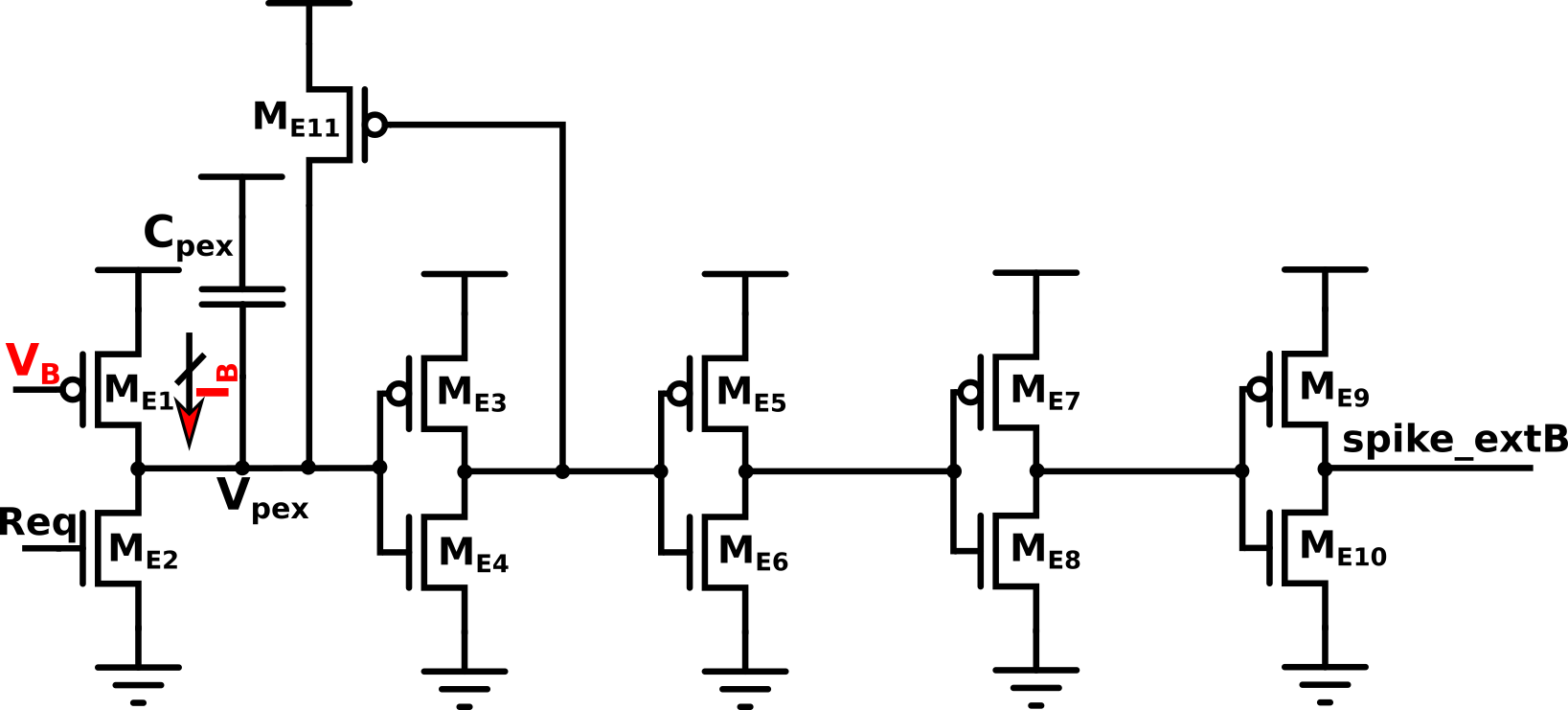}
    \caption{Pulse extender circuit schematic. Adapted from \cite{Rubino_etal19}.}
    \label{fig:PEX}
  \end{center}
\end{figure}

The \ac{ADEXP} \ac{FDSOI} circuit schematic can be subdivided into different functional blocks: 
An input \ac{DPI} ($M_{L1}$--$M_{L5}$)  models the neuron's leak conductance (LEAK).
A current-based positive feedback module ($M_{N1}$--$M_{N6}$) models the neuron's Sodium (Na+) activation and inactivation channels, and is coupled to a low-power current comparator (CC) block ($M_{C1}$--$M_{C9}$)  which triggers a spike as soon as the membrane current $I_{mem}$ exceeds the spiking threshold parameter $I_{thr}$.
A spike reset circuit with refractory period functionality ($M_{K1}$--$M_{K11}$) models the neuron's Potassium (K+) channels.
A negative feedback \ac{LPF} circuit implemented with an additional instance of a \ac{DPI}  ($M_{A1}$--$M_{A9}$) (AHP) emulates Calcium-dependent after-hyperpolarization Potassium currents observed in real neurons to produce the spike-frequency adaptation mechanism.
This circuit is driven each time the neuron produces an output spike event, which is conveyed to a pulse extender circuit (Fig.~\ref{fig:PEX}) to lengthen the duration of the spike-event and ensure proper sub-threshold operation of the \ac{DPI}.
Finally, an asynchronous digital handshaking (HS) block ($M_{HS1}$--$M_{HS4}$)  implements the interface to \ac{AER} circuits for transmitting the spikes as address-events to \ac{AER} routers and destinations.
This block generates the $Req$ and $Ack$ signals used to implement a four-phase handshaking cycle with the destination \ac{AER} circuits:  at rest, when the neuron current is below the spiking threshold, $Req$ and $Ack$ are both set to $gnd$.
As the $I_{mem}$ crosses the spiking threshold, provided $Ack$ is still at $gnd$, $Req$ is set to $V_{dd}$.
Once the \ac{AER} receiver consumes the event request and sets $Ack$ to $V_{dd}$ the neuron resets and $Req$ is pulled back to $gnd$.
As the \ac{AER} receiver senses this change, it should lower the $Ack$ signal, and the cycle can repeat.

To minimize leakage currents we reduced the Early effect of critical transistors by using a pseudo-cascode split-transistor sub-threshold technique~\cite{Qiao_etal17} (see transistors $M_{L4}$ and $M_{A5}$ of Fig.~\ref{fig:AdExpIF22nm}).
As for the \ac{DPI} synapse schematic of Fig.~\ref{fig:Syn22nm}, the ``$\times$2'' symbol in the figure denotes the presence of two transistors in series connected to form a diode-connected transistor. All transistors are configured with the default back-gate bias (0\,V).
All capacitors in this circuit are implemented using APMOM devices.
The value of the capacitances and their size is shown in Table~\ref{tab:Cap}.
\begin{table}
\caption{Capacitance values and sizes used in the neuron design}
\begin{center}
  \begin{tabular}{l*{20}{c}r}
\toprule
  &$C_{mem}$ &$C_{ahp}$&$C_{ref}$&$C_{pex}$&$C_{cc}$ \\
\midrule 
Value    &821\,fF&1\,pF& 102\,fF& 136\,fF& 116\,fF\\
Width &12\,$\mu$m&14\,$\mu$m&5.8\,$\mu$m&6\,$\mu$m&6\,$\mu$m\\
Length&12\,$\mu$m&14\,$\mu$m&5.5\,$\mu$m&7\,$\mu$m& 6\,$\mu$m\\
\bottomrule
  \end{tabular} 
  \label{tab:Cap}
\end{center}
\end{table}

The neuron behavior can be set by 6 tunable parameters (see bold labels in Fig.~\ref{fig:AdExpIF22nm}) that control the neuron's time constant ($V_{leak}$), its spike-frequency adaptation properties ($V_{a}$ and $V_{\tau\_ahp}$), its refractory period ($V_{ref}$), and its spiking threshold ($V_{thr}$). As the \ac{DPI} circuits of the LEAK and AHP blocks have been configured to have a gain term {$I_{gain}/I_{\tau}=1$} (similar to how the gain term was set to $4$ in the synapse circuit), the $V_{leak}$ and $V_{\tau\_ahp}$ signals can be tuned by modifying the $I_{gain}$ and $I_{gain\_ahp}$ currents respectively.

The spiking threshold parameter $V_{thr}$ controls the current comparator block  ($M_{C1}$--$M_{C9}$).
This is a novel circuit, modified from the one originally proposed in~\cite{Qiao_Indiveri16} to reduce the neuron's static and dynamic power consumption.
It has been introduced to decouple the slow and gradual changes of the neuron's membrane potential dynamics (represented by the $I_{mem}$ current) from the digital switching mechanism required to generate a spike.
This is necessary to minimize the switching time of the digital circuits, during which they can dissipate large amounts of power.
The $V_{inCC}$ voltage of the comparator circuit is set by the competition between the currents representing the spiking threshold, $I_{thr}$ , and $I_{mem}$.
If $I_{mem}$ is smaller than $I_{thr}$, this node is actively clamped to $V_{dd}$.
However as $I_{mem}$  approaches $I_{thr}$, $V_{inCC}$ drops sharply to produce a spike event with very low dynamic power consumption figures.
The transistor $M_{C3}$ was introduced in the CC block to reduce the neuron's power consumption in its resting (OFF) state: when $I_{in}$=0, the node $V_{inCC}$ is driven to $V_{dd}$ by $M_{C2}$ which turns on a discharging path to $gnd$ via $M_{C4}$. As both $M_{C2}$ and $M_{C4}$ are conducting, there exists a direct path between $V_{dd}$ and $gnd$ which undesirably adds to the static power consumption figure. The addition of $M_{C3}$ to this circuit breaks the discharging path to $gnd$, further reducing the neuron static power consumption.
%$M_{C3}$ is switched ON in parallel with the neuron, therefore it does not distort the spike generation.}
The $C_{CC}$ capacitance ensures that at each input current threshold crossing corresponds only one spike.

The pulse extender circuit depicted in Fig.~\ref{fig:PEX} is used to extend the spike-event pulse created by the neuron and to drive the spike-frequency adaptation circuit ($M_{A1}$--$M_{A9}$).
When a spike is produced and $Req$ goes to $V_{dd}$, the transistor $M_{E2}$ discharges the node $V_{pex}$ to $gnd$.
As a consequence, the node $spike_\_extB$ is discharged to $gnd$ switching ON $M_{A1}$ of Fig.~\ref{fig:AdExpIF22nm}.
The length of the  $spike_\_extB$ pulse is set by the $V_{B}$ bias voltage and the capacitance $C_{pex}$ . The higher the value of $V_{B}$, the slower $V_{pex}$ is charged back to $V_{dd}$ and hence the larger the extension. 
\section{Results}
\label{sec:results}

\subsection{Synaptic circuit simulation}

\begin{figure*}
  \centering
    \begin{subfigure}[]{0.45\textwidth}
      \includegraphics[width=\textwidth]{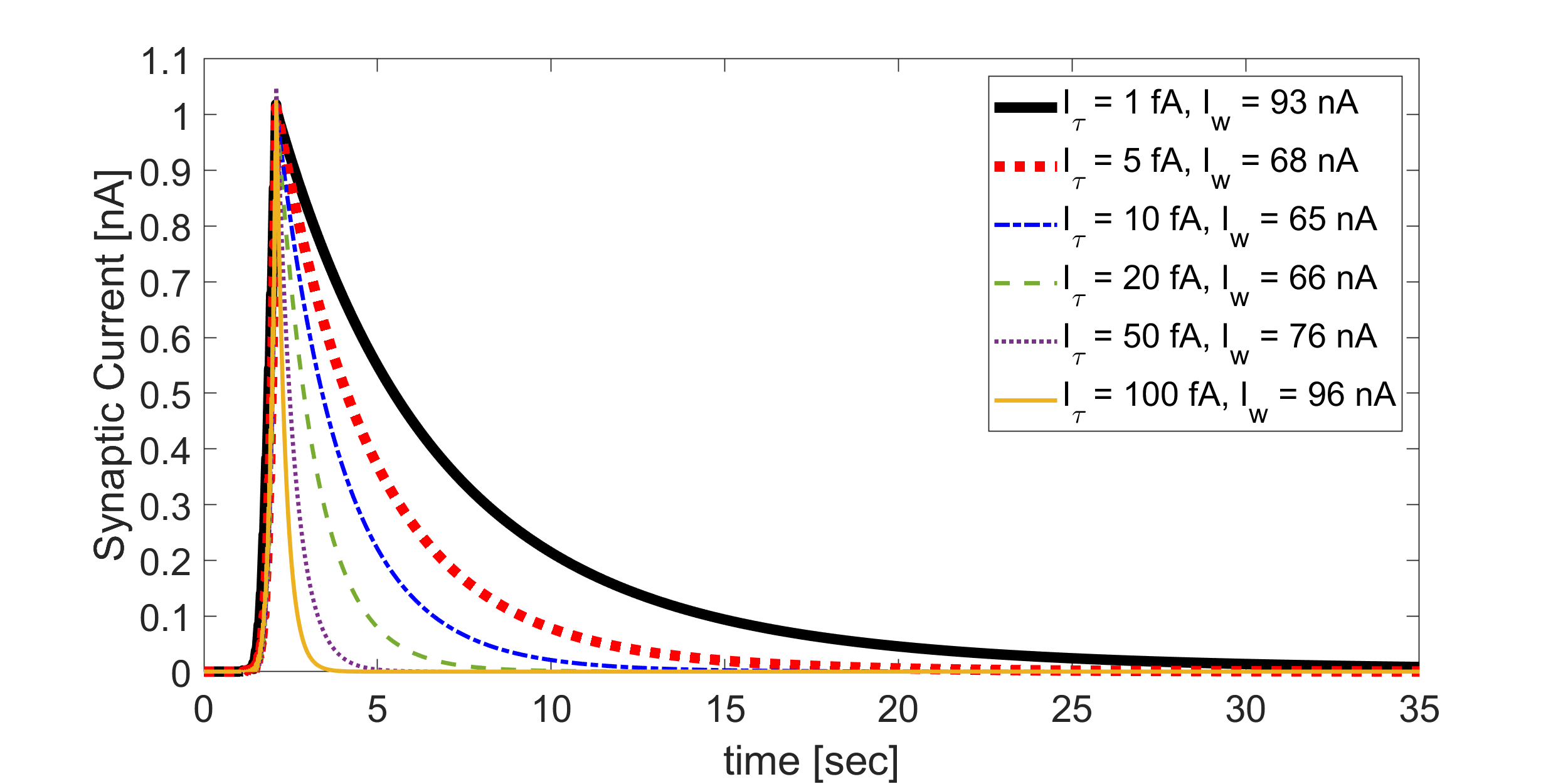}
      \caption{}
      \label{fig:dpi-slow}
    \end{subfigure}
    \begin{subfigure}[]{0.45\textwidth}
      \includegraphics[width=\textwidth]{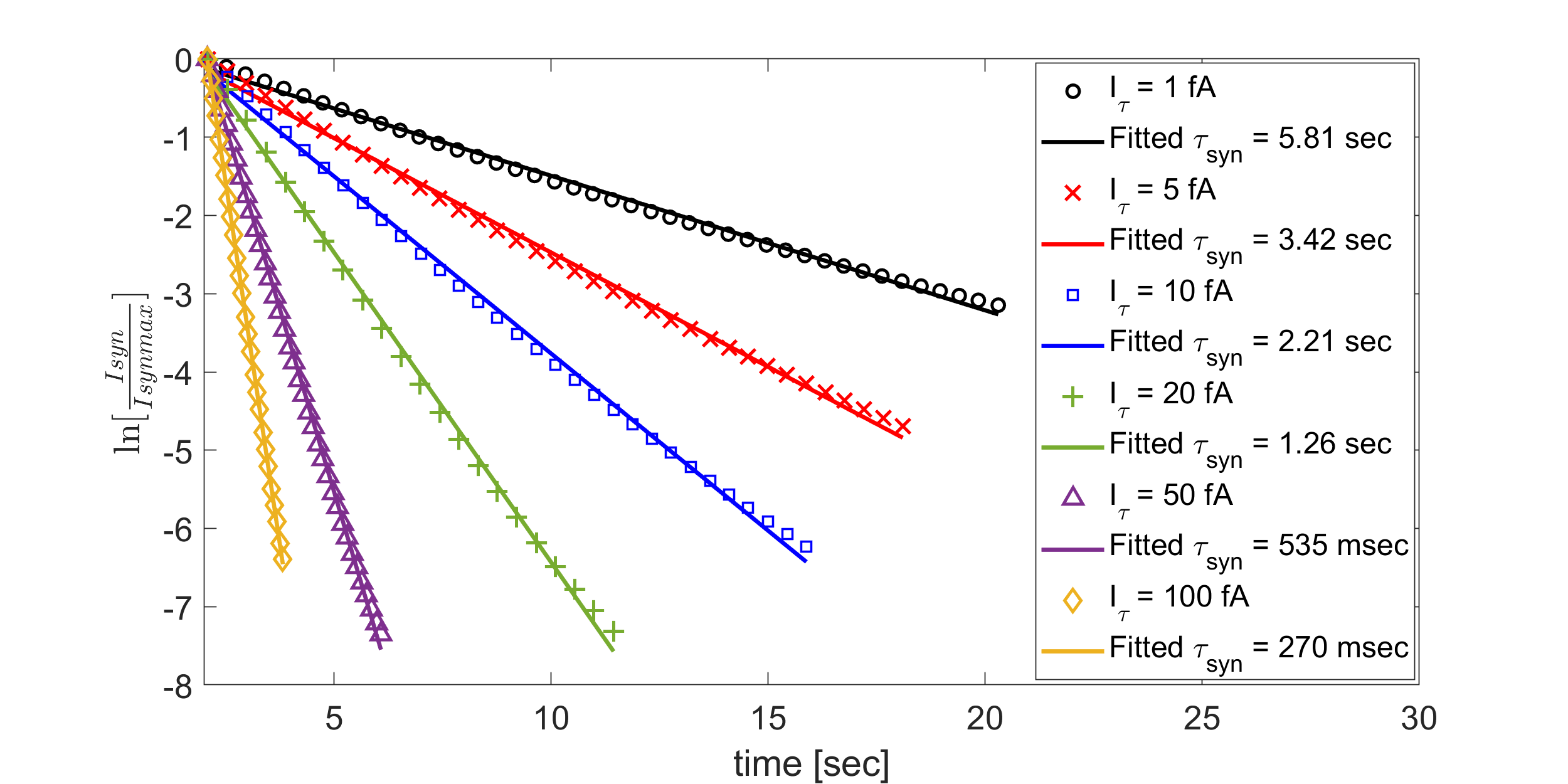}
      \caption{}
      \label{fig:dpi-slow-fit}
    \end{subfigure} \\
    \begin{subfigure}[]{0.45\textwidth}
      \includegraphics[width=\textwidth]{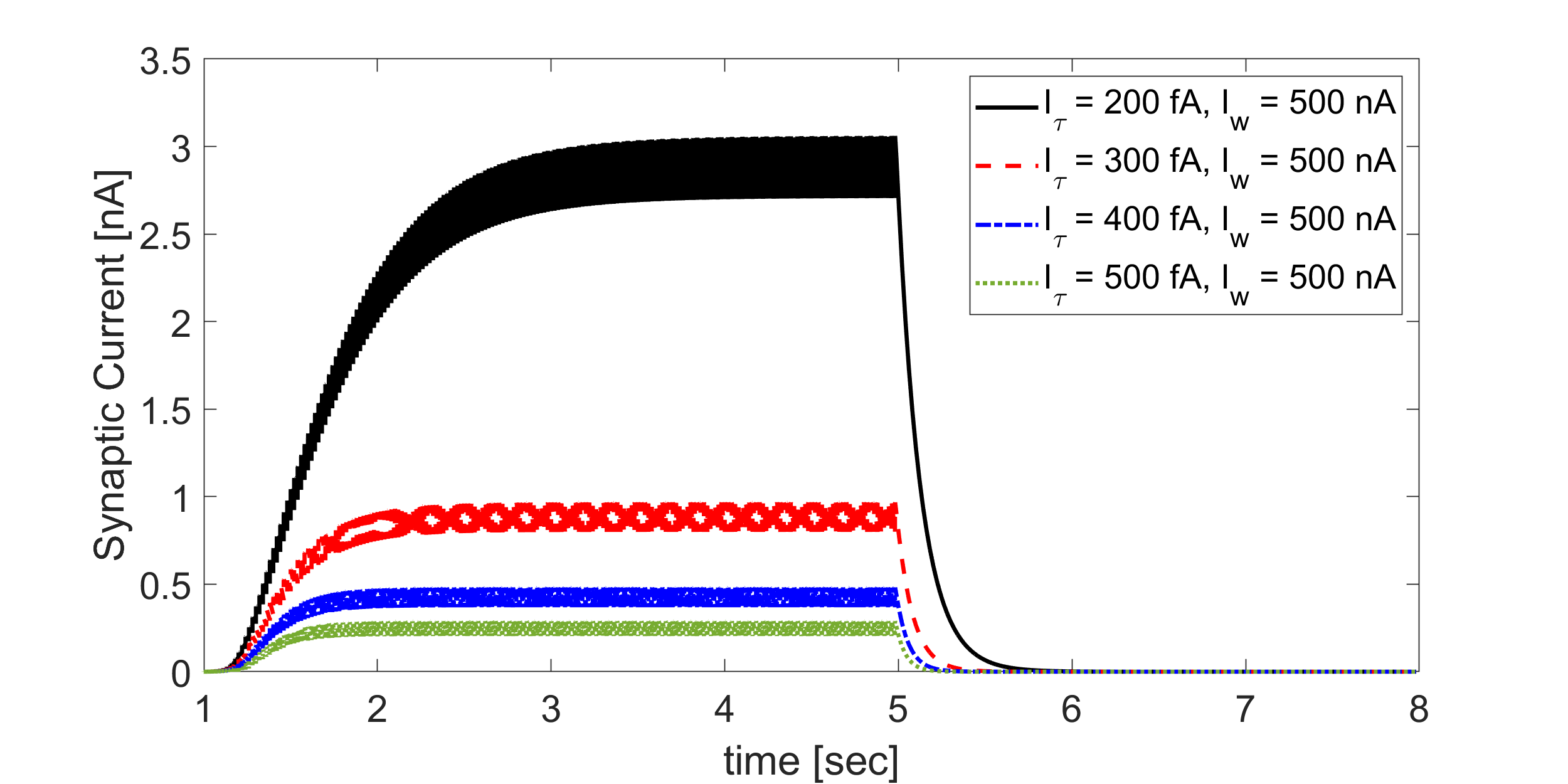}
      \caption{}
      \label{fig:dpi-fast}
    \end{subfigure}
    \begin{subfigure}[]{0.45\textwidth}
      \includegraphics[width=\textwidth]{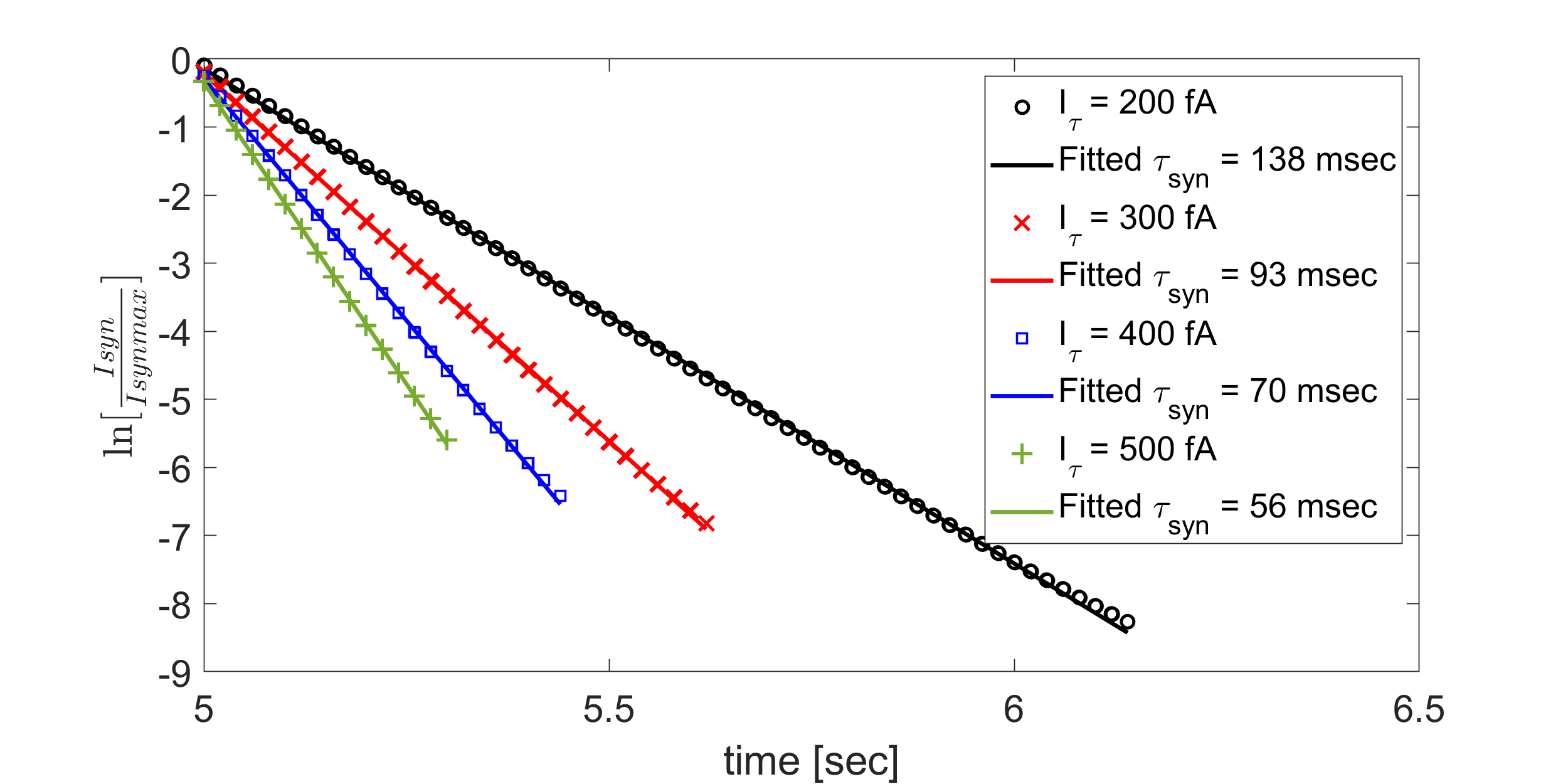}
      \caption{}
      \label{fig:dpi-fast-fit}
    \end{subfigure}
    \caption{Synapse $I_{syn}$ \ac{EPSC} measurements and fits with the natural logarithm of the normalized \ac{EPSC} for estimating its time constants $\tau_{syn}$: (a) \& (b) for a 50\,Hz stimulus applied for 1\,sec  with synaptic time constants in the range 250\,msec--6\,sec. (c) \& (d) for a 50\,Hz stimulus applied for 5\,sec, with synaptic time constants in the range 50\,msec--150\,msec, causing the synapse to reach different steady state levels.}
    \label{fig:DPIresult}
\end{figure*}

The simulations presented in this section demonstrate how the \ac{DPI} synapse presented in Section~\ref{sec:synapse} can achieve very large time constants and high synaptic efficacy. The results demonstrating the integration and steady-state profiles of the synapse \ac{EPSC} ($I_{syn}$) are shown in Fig.~\ref{fig:dpi-slow} and ~\ref{fig:dpi-fast} respectively. To assess the corresponding synaptic time constants ($\tau_{syn}$), we plot the linear fits to the natural logarithm of the normalized \ac{EPSC} in Fig.~\ref{fig:dpi-slow-fit} and~\ref{fig:dpi-fast-fit}.

To measure the circuit's dynamic range we stimulate it with a pulse train of 50\,Hz rate, with each pulse lasting 100\,nsec; and we sweep $I_{\tau}$ to cover a wide spectrum of time constants extending from 50\,msec to 6\,sec. To demonstrate the integration property of the synapse at lower $I_{\tau}$ values (i.e., larger time constants), we apply the pulse train for 1\,sec, adjust the weight $I_{w}$ accordingly to obtain a peak integrated response of 1\,nA, and measure the decay time of $I_{syn}$. 
Moreover, to observe the different steady-state response behavior, we extend the pulse train stimulation duration to 5\,sec, fix $I_w$ to 500\,nA and use smaller time constant values which are comparable with the input pulse train inter-spike interval.

Figure \ref{fig:dpi-slow} shows the synapse response in integration mode with the time constants in the range 250\,msec--6\,sec. Here, the inter-spike interval of 20\,msec is much smaller than the synaptic time constants, and thus the \ac{EPSC} charges up to 1\,nA for all $I_{\tau}$ values. However, to obtain equal peak \ac{EPSC} magnitudes we had to make small adjustments of $I_w$ for the values of $I_{\tau}$ between 5 and 50\,fA, and larger adjustments for $I_{\tau}$=1\,fA and $I_{\tau}$=100\,fA.
As the value of $I_{\tau}$ decreases to fA values, the \ac{APMOM} capacitor parasitic effects become non-negligible: specifically, the capacitor's leak increases the effective value of $I_{\tau}$. For ideal values of $I_{\tau}$=1\,fA, the effective synaptic efficacy term $\frac{I_{gain}}{I_{\tau}}$ is in practice much less than the nominal value of 4. To compensate for this effect it is therefore necessary to increase the value of $I_w$ (e.g., see solid line in Fig.~\ref{fig:dpi-slow}).

%The larger weight increase for $I_{\tau}$=1\,fA case is needed to compensate for the reduction in the synaptic efficacy due to the \ac{APMOM} capacitor parasitic effects. \MP{Essentially, the capacitor's leak increases the effective $I_{\tau}$ which has a significant effect on the synaptic efficacy when $I_{\tau}$ is set to very small values (e.g., 1\,fA for $\tau_{syn=}$5.81\,s).}  \CL{I would prefer the old version commented below.} 
%Although the synaptic efficacy term $\frac{I_{gain}}{I_{\tau}}$ of Eq~\eqref{eq:dpi_syn} is set to $4$, for very long time constants (e.g., $\tau_{syn=}$5.81\,s), the \ac{APMOM} capacitor's parasitic reduces it significantly, because the capacitor leakage can strongly affect the ultra-low values of $I_{\tau}$ while $I_{gain}$ remains constant.

Figure \ref{fig:dpi-fast} shows how the synapse reaches a steady-state with shorter time constants and a longer stimulus duration. In this mode of operation, the discharge of the synapse during the inter-spike interval balances out with the charge induced by the spikes. Hence, the steady-state value scales with the time constant. In addition, the \ac{EPSC} at steady-state features fluctuations around its peak value due to the ongoing rapid synaptic charging-discharging process. 

The fits to the synaptic current data illustrated in Fig.~\ref{fig:dpi-slow-fit} and~\ref{fig:dpi-fast-fit} verify that the synapse behaves as a first order system as Eq.~\eqref{eq:dpi_syn} suggests. Table~\ref{tab:tau} compares the theoretical time constants to the values obtained by linear fitting. 
The table shows that the fitting results are in very good agreement with the theory for fast synaptic dynamics while the difference increases as the synapse becomes slower. This is due to the \ac{APMOM} capacitor leakage building on $I_{\tau}$ which limits the maximum time constant of the circuit. This limitation is particularly significant for $I_{\tau}$=1\,fA, where the capacitor leakage dominates the synaptic discharging process. As a consequence, the time constant saturates at $\tau_{syn}$=5.81\,sec above which the circuit cannot extend. Based on the results, the capacitor leakage is found to constitute a 3--4\,fA of peak baseline current which is much higher than the cumulative leakage of all transistors. Although the reduction in the transistor leakage has increased the available time constant range, the capacitive leakage would need to be reduced to sub-femto-Ampere regime as well, in order to increase the time constants further up to 30\,sec as the theory suggests. Increasing the capacitance size is not a viable option, since the capacitor leakage scales up with its area as well. Moreover, this capacitive leakage is voltage dependent hence utilizing different synaptic weights or pulse configurations can result in slight deviations in the available time constant range. 

\begin{table}
  \caption{Comparison between the measured and theoretical time constants $\tau_{syn}$ of the \ac{DPI} synapse for $C_{syn}$=821\,fF, $\kappa$=0.75, and $U_T$=25\,mV}
  \centering
  \begin{tabular}{l*{6}{c c || c c c}}
    \toprule
    $I_{\tau}$ & Theoretical  & Linear Fit  & $I_{\tau}$  & Theoretical & Linear Fit \\
    \midrule 
    1\,fA & 27.37\,sec  &  5.81\,sec  & 100\,fA & 274\,msec & 270\,msec \\
    5\,fA & 5.47\,sec  &  3.42\,sec  & 200\,fA & 137\,msec & 138\,msec \\
    10\,fA & 2.74\,sec  &  2.21\,sec  & 300\,fA & 91\,msec & 93\,msec \\
    20\,fA & 1.37\,sec  &  1.26\,sec  & 400\,fA & 68\,msec & 70\,msec \\
    50\,fA & 547\,msec  &  535\,msec  & 500\,fA & 55\,msec & 56\,msec \\
    \bottomrule
  \end{tabular}
  \label{tab:tau}
\end{table}

In general, the designed \ac{DPI} synapse circuit with ultra-low leakage capability can offer a wide dynamic range of time constants with sub-pico-Ampere $I_{\tau}$ values and can generate \ac{EPSC} on the order of nano-Amperes for 100\,nsec of pulse duration. The ultra-low current operation of the circuit makes time constants up to several seconds achievable with capacitors below 1\,pF which reduces the layout area and enables denser integration in addition to the benefit of reducing the power consumption of the overall circuit. 

\begin{figure*}
  \begin{center}
    \begin{subfigure}[]{0.325\textwidth}
      \includegraphics[width=\textwidth]{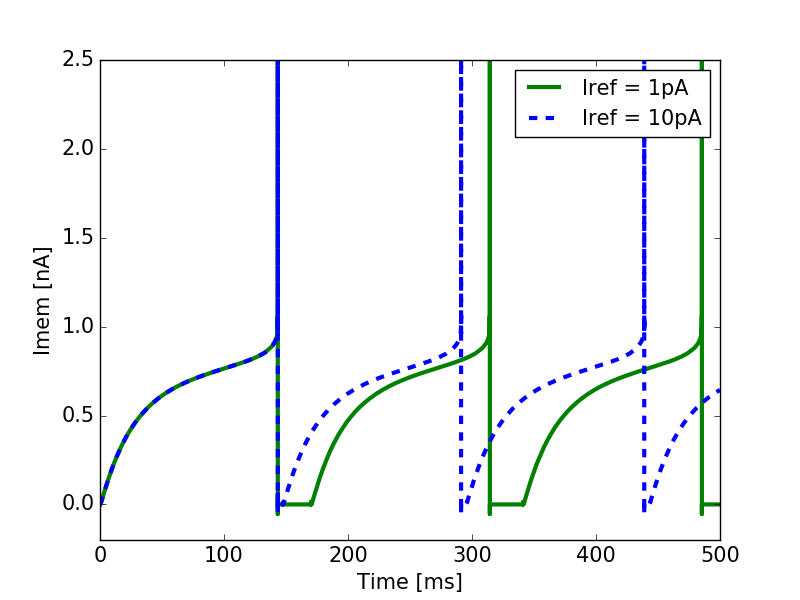}
      \caption{}
      \label{fig:ref}
    \end{subfigure}
    \begin{subfigure}[]{0.325\textwidth}
      \includegraphics[width=\textwidth]{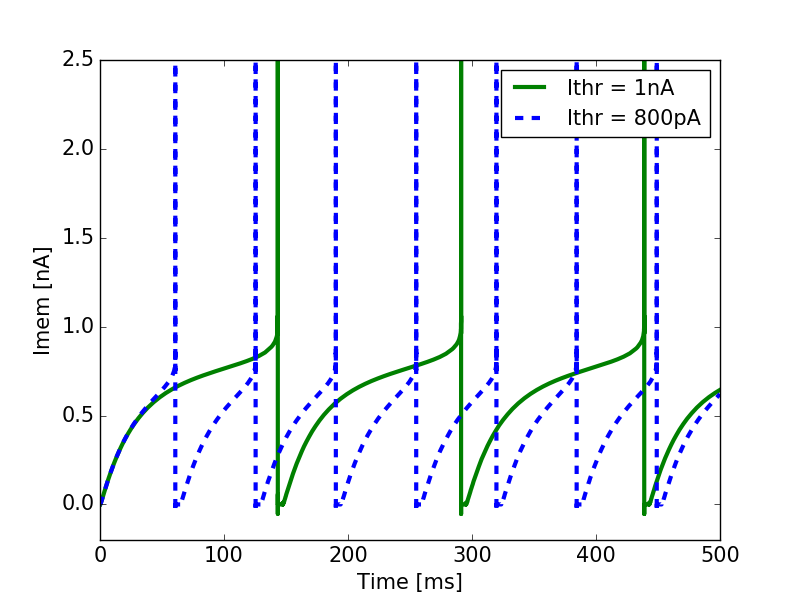}
      \caption{}
      \label{fig:thr1}
    \end{subfigure}
    \begin{subfigure}[]{0.325\textwidth}
      \includegraphics[width=\textwidth]{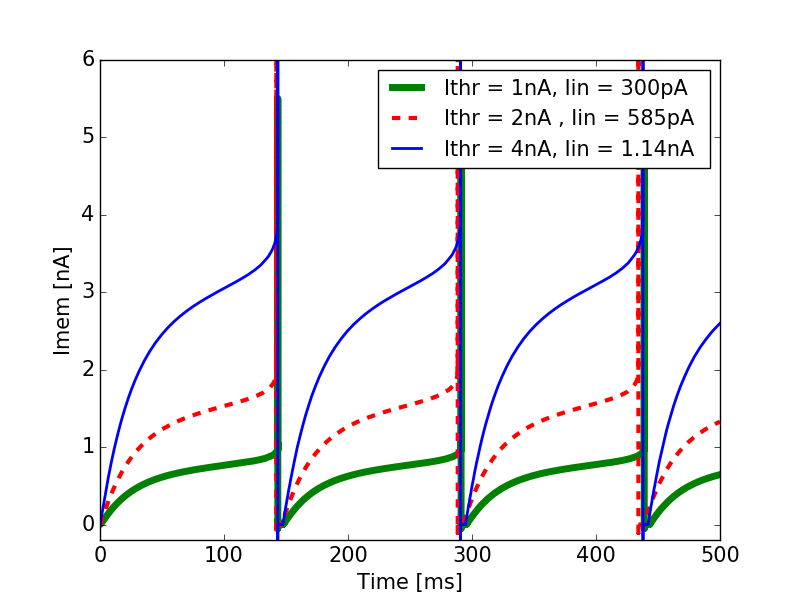}
      \caption{}
      \label{fig:thr2}
    \end{subfigure}
    \caption{\ac{FDSOI} neuron biologically plausible behaviour: (a) Membrane current $I_{mem}$ for two different values of $I_{ref}$, (b) $I_{mem}$ for two different values of $I_{thr}$ keeping the same $I_{in}$, (c) $I_{mem}$ for three different values of $I_{thr}$ changing $I_{in}$ to obtain the same spiking frequency.}
    \label{fig:neuron_dyn}
  \end{center}
\end{figure*}

\subsection{Neuron circuit simulation}
\label{sec:neur-circ-simul}

In this section we present the \ac{FDSOI} neuron simulation results, demonstrating examples of biologically plausible behaviors, characterizing its power consumption properties, and quantifying the effects of device mismatch on its response properties.
Figure~\ref{fig:neuron_dyn} shows the response of the neuron to a step input current for different parameter settings: Fig.~\ref{fig:ref} shows the membrane current $I_{mem}$ for two different values of $I_{ref}$. As the $I_{ref}$ increases, the refractory period is shorter and hence the neuron's maximum spiking frequency increases.  Figure~\ref{fig:thr1} and~\ref{fig:thr2} show the $I_{mem}$ for different values of $I_{thr}$ when keeping the same $I_{in}$ (Fig.~\ref{fig:thr1}) and when changing $I_{in}$ (Fig.~\ref{fig:thr2}) to obtain the same spiking frequency as $I_{thr}$ changes. As $I_{thr}$ increases the neuron is less facilitated to spike as it has to integrate more current to reach the threshold. Higher $I_{thr}$ leads to more time needed in the integrating phase and therefore less frequent spikes, which gives lower spiking frequency.
\begin{figure*}
  \begin{center}
    \begin{subfigure}[]{0.325\textwidth}
      \includegraphics[width=\textwidth]{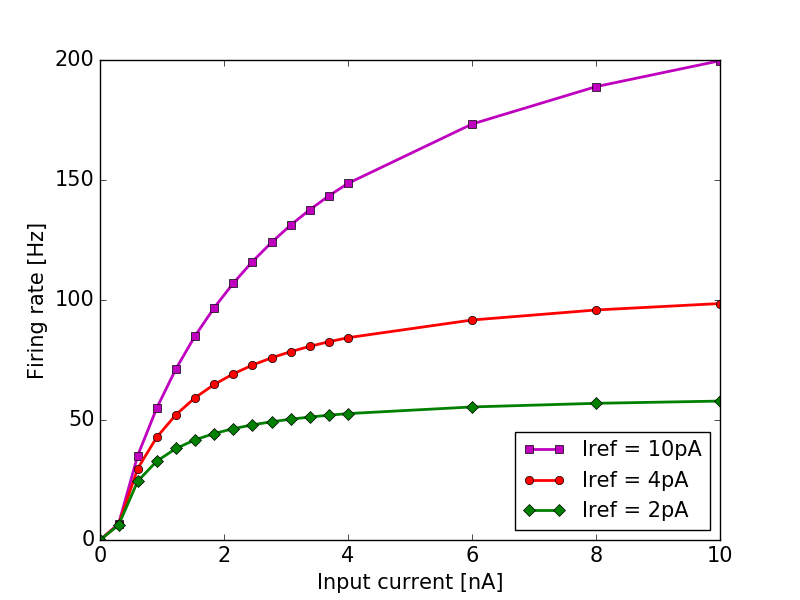}
      \caption{}
      \label{fig:ref2}
    \end{subfigure}
    \begin{subfigure}[]{0.325\textwidth}
      \includegraphics[width=\textwidth]{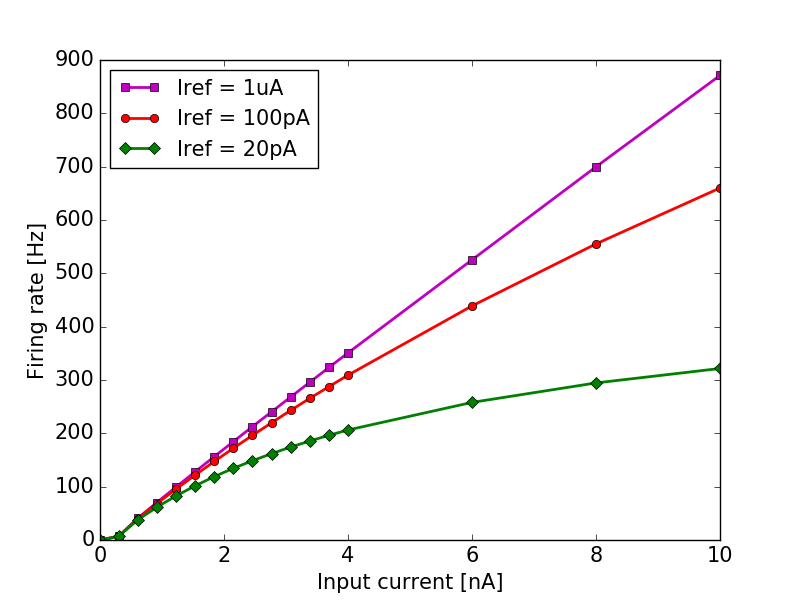}
      \caption{}
      \label{fig:ref3}
    \end{subfigure}
    \begin{subfigure}[]{0.325\textwidth}
      \includegraphics[width=\textwidth]{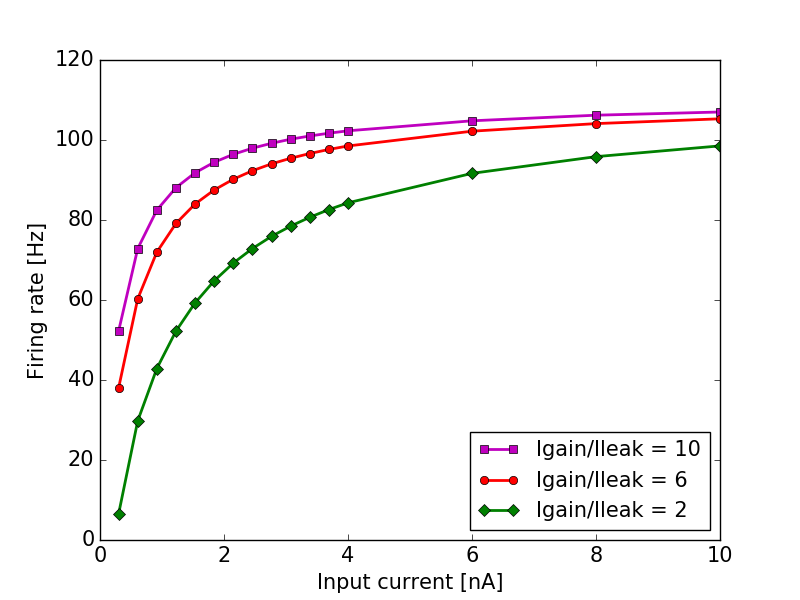}
      \caption{}
      \label{fig:gain}
    \end{subfigure}
    \caption{Firing rate vs Input current (F-I) curve: (a) for lower values of $I_{ref}$ hence longer refractory period, (b) for higher value of $I_{ref}$ hence shorter refractory period and (c) for different values of $I_{gain}$/$I_{leak}$ ratio. Adapted from \cite{Rubino_etal19}.}
    \label{fig:neuron_dyn_2}
  \end{center}
\end{figure*}

Figure~\ref{fig:ref2} and~\ref{fig:ref3} show the neurons F-I curve for different $I_{ref}$ bias settings. The neurons average firing rate increases linearly with the input current, until it reaches a saturation level that depends on the refractory period setting (see Fig.~\ref{fig:ref2}). The saturation frequency depends on the duration of the refractory period, as the $I_{ref}$ increases the refractory period is shorter and hence the neuron's maximum spiking frequency increases. For shorter refractory periods  (Fig.~\ref{fig:ref3}) the maximum spiking frequency is linear with the input current for a larger range of $I_{in}$ and it reaches saturation only for very high values of $I_{in}$. When $I_{ref}$ is 1\,$\mu$A, hence when the refractory period is very short (few hundreds of ns), the maximum spiking frequency does not saturate for the chosen $I_{in}$ range.

Figure~\ref{fig:gain} shows the neuron spiking frequency versus input current (F-I curve), for different settings of the $I_{gain}$/$I_{leak}$ bias ratio. We modify the default ratio ($I_{gain}$/$I_{leak}$=1) to higher values. As expected, increase in the ratio results in the increase of the neuron's firing rate.
\begin{figure}
  \centering
    \includegraphics[width=0.4\textwidth]{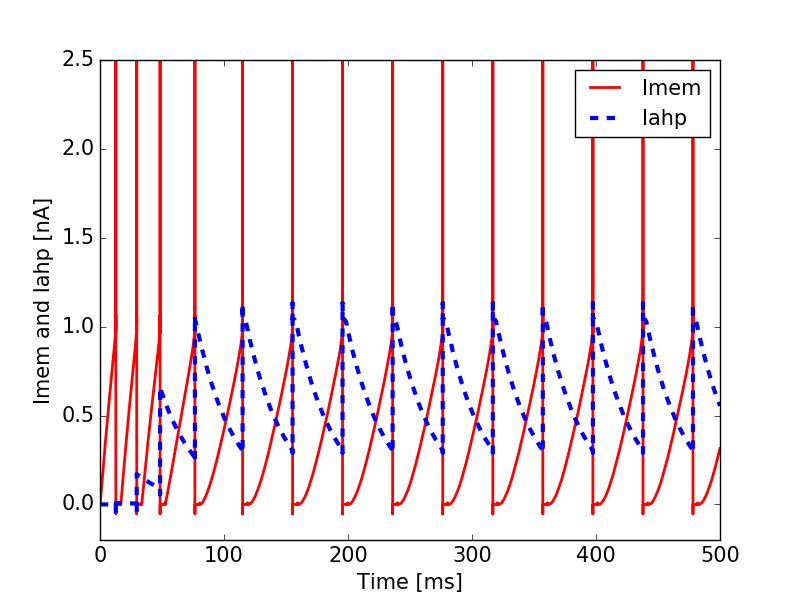}
    \caption{Spike-frequency adaptation: Membrane current and after-hyperpolarization current trace over time. Adapted from \cite{Rubino_etal19}.}
    \label{fig:Iahp}
\end{figure}

Figure~\ref{fig:Iahp} demonstrates the spike-frequency adaptation behavior, obtained by appropriately tuning the relevant parameters in the AHP block of Fig.~\ref{fig:AdExpIF22nm} and measuring the neuron's step response to a constant injection current. At each spike the $I_{ahp}$ current increases, decreasing the total current charging $C_{mem}$. When $I_{ahp}$ reaches the steady state, the neuron spiking frequency remains constant to a lower value compared to the initial one.

\subsection{Energy per spike}
\label{sec:e_spike}

\begin{figure}
  \centering
  \includegraphics[width=0.4\textwidth]{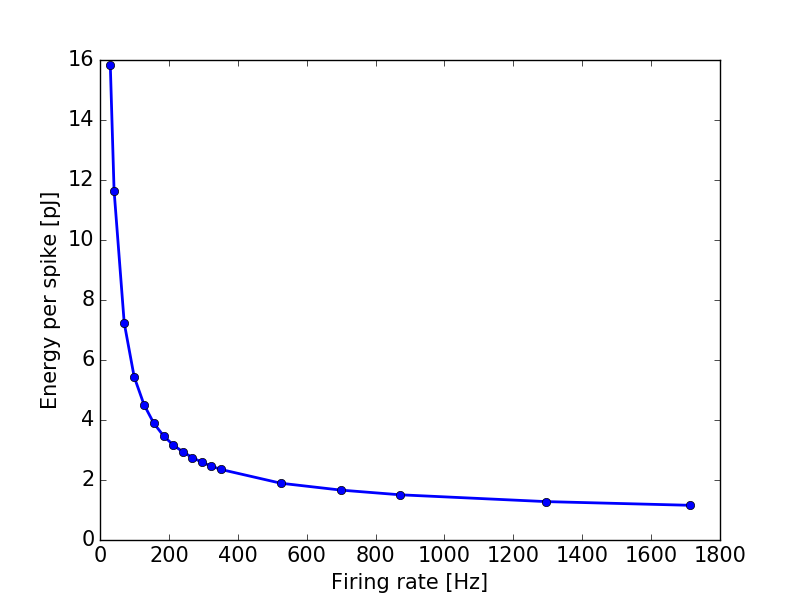}
  \label{fig:eperSfreq}
  \caption{Energy per spike estimation: Energy per spike vs Firing rate.}
  \label{fig:eperS}
\end{figure}

Once proven that the design is able to reproduce a biologically plausible behavior, we evaluated whether it can implement massively parallel large-scale neuromorphic processors.
The energy per spike is equal to:
\begin{equation}
  \frac{Energy}{Freq \cdot Time} =  \frac{Power \cdot Time}{Freq \cdot Time} = \frac{Power}{Freq}
\label{eq:eperS}
\end{equation}
%\caption{Energy per spike estimation: $Energy$ the total energy consumed, $Power$ total power consumed, $Time$ simulation time and $Freq$ the maximum spiking frequency.}
where $Energy$, the total energy consumed, is the product of total power consumed ($Power$) and simulation time ($Time$) and $Freq$ is the maximum spiking frequency.
As is shown in Fig.~\ref{fig:eperS}, the energy per spike for lower frequencies is in the order of tens of pJ, while it decreases to 1\,pJ for higher frequencies. 
This is due to the fact that in lower frequencies the inverters spend more time in their high gain region with both transistors conducting, since the $V_{mem}$ at their input is charging much slower compared to the case with higher frequencies.

We compare the energy per spike of our proposed neuron with previously published state-of-the-art neuromorphic processors in Table ~\ref{tab:Prev}.
\begin{table}
  \caption{Energy per spike comparison with previous works}
  \centering
  \begin{tabular}{l*{5}{c}}
    \toprule
    Work  & \cite{Moradi_etal18} & \cite{Mayr_etal16}&\cite{Qiao_Indiveri16}& \textbf{This work} \\
    \midrule 
    Techn. &180\,nm &28\,nm& 28\,nm&22\,nm \\
    &CMOS &CMOS& FDSOI&FDSOI \\
    Type         & Mixed & Mixed & Mixed& Mixed \\
    $V_{dd}$        & 1.8\,V & 0.7-1\,V &  1\,V& 0.8\,V \\
    Freq & 30\,Hz  & - & 30\,Hz & 30\,Hz\\
    En./spike  & 883\,pJ  & 2.3\,nJ-30\,nJ & 50\,pJ & 16\,pJ \\
    \bottomrule
  \end{tabular}
  \label{tab:Prev}
\end{table}

The neuron designed in this work consumes less energy per spike compared to a similar circuit \cite{Qiao_Indiveri16} in a similar technology (28\,nm \ac{FDSOI}) at a biological plausible spiking frequency (30\,Hz). Moreover, the circuits used in \cite{Qiao_Indiveri16} and our work have similar $V_{dd}$ and $C_{mem}$, which we can consider as the predominant capacitance for power consumption. Therefore, according to the scaling factor, the energy per spike of the circuit proposed in~\cite{Qiao_Indiveri16} will be similar when scaled to the 22\,nm process. Hence the differences reported here can be explained by the optimizations made at the circuit design level.

The neuron circuit energy consumption at higher frequencies is compared with the Sigma-Delta neuron proposed in~\cite{Nair_Indiveri19}, which is one of the most recent mixed-signal silicon neuron circuit designs presented in the literature.
Since the Sigma-Delta neuron presented in~\cite{Nair_Indiveri19} was optimized for operation at higher frequencies in a range of 1\,kHz to 10\,MHz, we compare the energy per spike between these circuits in these ranges: the neuron proposed in this work consumes 1\,pJ@2.1\,kHz, approximately one order of magnitude less  than the Sigma-Delta neuron (10\,pJ). %This data point (2.1\,kHz) is not shown in Fig.~\ref{fig:eperS} for making our interested range of operation more visible.

\begin{figure*}
  \centering
  \includegraphics[width=0.8\textwidth]{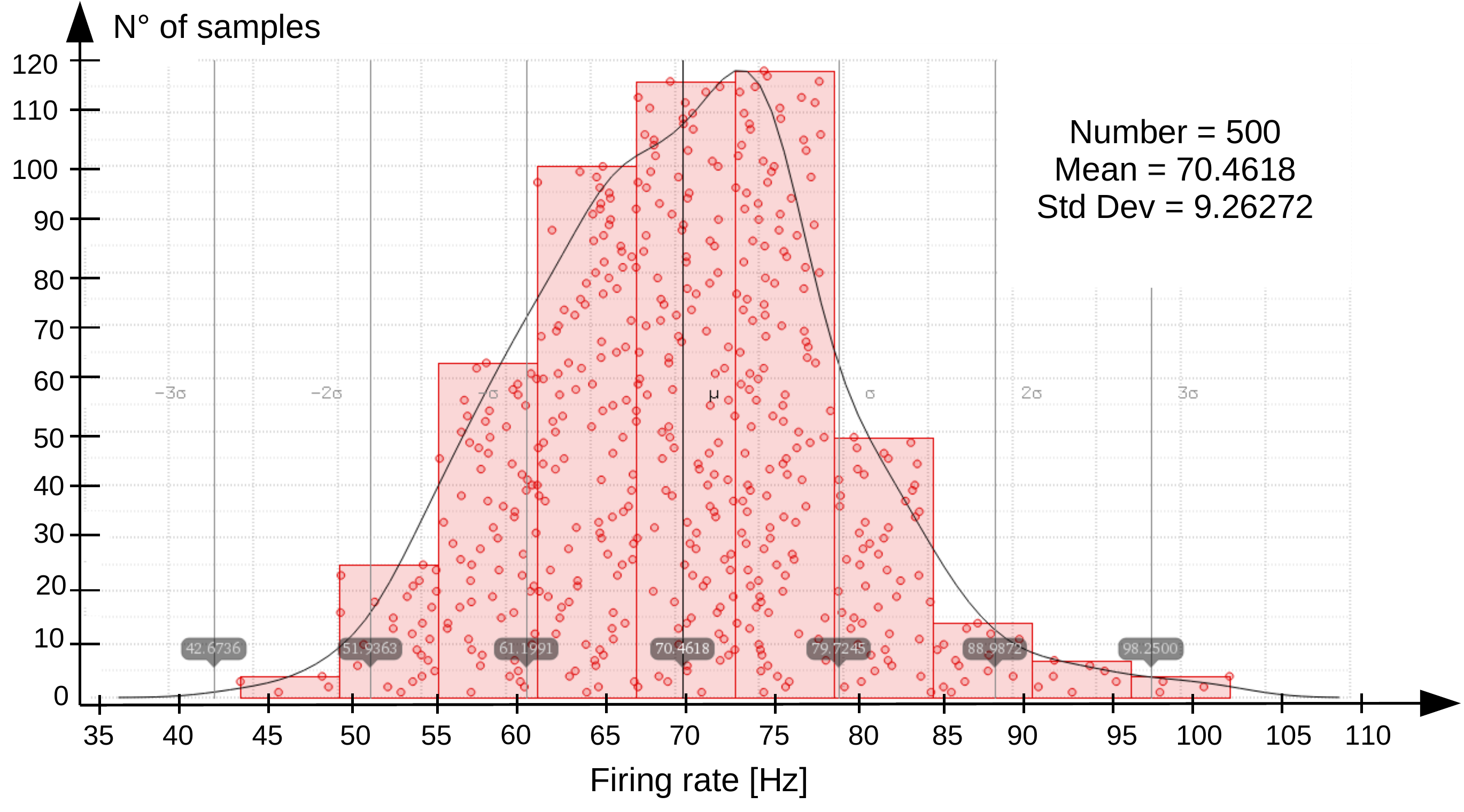}
  \caption{Monte Carlo analysis result distribution of the neuron circuit. Adapted from \cite{Rubino_etal19}.}
  \label{fig:mcs}
\end{figure*}

\subsection{Monte Carlo Analysis}
\label{sec:MCS}
To evaluate the sensitivity of the circuit to device mismatch we ran a series of Monte Carlo simulations. We performed this analysis with 500 runs for this neuron circuit, with DC current injected in the LEAK block in Fig.~\ref{fig:AdExpIF22nm}, and with bias currents set to obtain a firing rate of approximately 70\,Hz while switching off the spike-frequency adaptation circuit.

The mean of the distribution obtained is centered around the expected value ($\approx$70\,Hz) and the standard deviation is equal to 9.26 (see Fig.~\ref{fig:mcs}). The variability of the neuron circuit is thus 13 \%.
Our analysis found that this variability is governed by the LEAK block and the first part of the CC, where the comparison between $I_{mem}$ and $I_{thr}$ is made ($M_{C1}$ and $M_{C2}$). In particular, in the LEAK block the transistors more affected by process variation are $M_{L2}$ and $M_{L3}$.
The K+ block, in particular $M_{K4}$, also shows sensitivity to device mismatch, but it is negligible compared to the other two blocks.

\section{Conclusion} 
\label{sec:concl}   
We determined process and circuit parameters in order to implement efficient (low power and slow dynamics) analog neuron circuits using an advanced scaled 22\,nm \ac{FDSOI} process.
We optimized the design of the synapse and neuron circuits for producing biologically plausible neural dynamics, with time constants matched to those of natural signals, such as speech or bio-signals.

The presented silicon synapse circuit can achieve time constants of up to 6\,sec without having to increase the synaptic capacitance $C_{syn}$ over 1\,pF.
The neuron circuit presented has an energy per spike of tens of pJ for lower frequencies and pJ for higher frequencies, which is considerably lower compared to an analogous neuron design implemented in a 180\,nm CMOS process~\cite{Moradi_etal18}. 
Furthermore, it consumes less compared to a more recent design \cite{Qiao_Indiveri16} at biologically plausible frequencies and it consumes one order of magnitude less compared to the state-of-the-art neuron circuit~\cite{Nair_Indiveri19} at higher frequencies.
We studied the mismatch sensitivity of the neuron circuit by performing Monte Carlo simulations and identified the parts of the circuit that are most critical to be optimized for variations, showing how the more sensitive sub-parts of the silicon neuron circuit are the LEAK block and the first part of the CC block.
In summary in this paper we demonstrate how it is possible to exploit the features of advanced 22\,nm FDSOI processes to design complex analog circuits that can be used to implement low-power neuromorphic processors for edge computing sensory-processing tasks and, more generally, ``neuromorphic intelligence'' applications.

\section*{Acknowledgment}
We are grateful to Mohammad Ali Sharif Shazileh, Manu Nair, and Elisa Donati for fruitful discussions on the manuscript and circuit design. 

\color{black}
\bibliographystyle{IEEEtran}
\bibliography{biblioncs}

\begin{thebibliography}{10}
\providecommand{\url}[1]{#1}
\csname url@samestyle\endcsname
\providecommand{\newblock}{\relax}
\providecommand{\bibinfo}[2]{#2}
\providecommand{\BIBentrySTDinterwordspacing}{\spaceskip=0pt\relax}
\providecommand{\BIBentryALTinterwordstretchfactor}{4}
\providecommand{\BIBentryALTinterwordspacing}{\spaceskip=\fontdimen2\font plus
\BIBentryALTinterwordstretchfactor\fontdimen3\font minus
  \fontdimen4\font\relax}
\providecommand{\BIBforeignlanguage}[2]{{%
\expandafter\ifx\csname l@#1\endcsname\relax
\typeout{** WARNING: IEEEtran.bst: No hyphenation pattern has been}%
\typeout{** loaded for the language `#1'. Using the pattern for}%
\typeout{** the default language instead.}%
\else
\language=\csname l@#1\endcsname
\fi
#2}}
\providecommand{\BIBdecl}{\relax}
\BIBdecl

\bibitem{Davari_etal95}
B.~Davari, R.~Dennard, and G.~Shahidi, ``{CMOS} scaling for high performance
  and low power -- the next ten years,'' \emph{Proceedings of the {IEEE}},
  vol.~83, no.~4, pp. 595--606, April 1995.

\bibitem{Backus78}
\BIBentryALTinterwordspacing
J.~Backus, ``Can programming be liberated from the {von Neumann} style?: a
  functional style and its algebra of programs,'' \emph{Communications of the
  ACM}, vol.~21, no.~8, pp. 613--641, 1978. [Online]. Available:
  \url{http://doi.acm.org/10.1145/359576.359579}
\BIBentrySTDinterwordspacing

\bibitem{Indiveri_Liu15}
G.~Indiveri and S.-C. Liu, ``Memory and information processing in neuromorphic
  systems,'' \emph{Proceedings of the {IEEE}}, vol. 103, no.~8, pp. 1379--1397,
  2015.

\bibitem{Maass_Markram04}
W.~Maass and H.~Markram, ``On the computational power of circuits of spiking
  neurons,'' \emph{Journal of computer and system sciences}, vol.~69, no.~4,
  pp. 593--616, 2004.

\bibitem{Zambrano_Bohte16}
D.~Zambrano and S.~M. Bohte, ``Fast and efficient asynchronous neural
  computation with adapting spiking neural networks,'' \emph{arXiv e-prints},
  2016.

\bibitem{Neftci18}
\BIBentryALTinterwordspacing
E.~O. Neftci, ``Data and power efficient intelligence with neuromorphic
  learning machines,'' \emph{iScience}, vol.~5, pp. 52--68, 2018. [Online].
  Available:
  \url{http://www.sciencedirect.com/science/article/pii/S2589004218300865}
\BIBentrySTDinterwordspacing

\bibitem{Lukosevicius_Jaeger09}
M.~Luko{\v{s}}evi{\v{c}}ius and H.~Jaeger, ``Reservoir computing approaches to
  recurrent neural network training,'' \emph{Computer Science Review}, vol.~3,
  no.~3, pp. 127--149, 2009.

\bibitem{Bellec_etal18}
G.~Bellec, D.~Salaj, A.~Subramoney, R.~Legenstein, and W.~Maass, ``Long
  short-term memory and learning-to-learn in networks of spiking neurons,'' in
  \emph{Advances in Neural Information Processing Systems}, 2018, pp. 787--797.

\bibitem{Mead90}
C.~Mead, ``Neuromorphic electronic systems,'' \emph{Proceedings of the {IEEE}},
  vol.~78, no.~10, pp. 1629--36, 1990.

\bibitem{Chicca_etal14}
E.~Chicca, F.~Stefanini, C.~Bartolozzi, and G.~Indiveri, ``Neuromorphic
  electronic circuits for building autonomous cognitive systems,''
  \emph{Proceedings of the {IEEE}}, vol. 102, no.~9, pp. 1367--1388, 9 2014.

\bibitem{Indiveri_Sandamirskaya19}
G.~Indiveri and Y.~Sandamirskaya, ``The importance of space and time for signal
  processing in neuromorphic agents,'' \emph{{IEEE} Signal Processing
  Magazine}, vol.~36, no.~6, pp. 16--28, 2019.

\bibitem{Moradi_etal18}
S.~Moradi, N.~Qiao, F.~Stefanini, and G.~Indiveri, ``A scalable multicore
  architecture with heterogeneous memory structures for dynamic neuromorphic
  asynchronous processors ({DYNAPs}),'' \emph{Biomedical Circuits and Systems,
  {IEEE} Transactions on}, vol.~12, no.~1, pp. 106--122, Feb. 2018.

\bibitem{Bauer_etal19}
F.~Bauer, D.~Muir, and G.~Indiveri, ``Real-time ultra-low power {ECG} anomaly
  detection using an event-driven neuromorphic processor,'' \emph{Biomedical
  Circuits and Systems, {IEEE} Transactions on}, 2019, (in press).

\bibitem{Rubino_etal19}
A.~Rubino, M.~Payvand, and G.~Indiveri, ``Ultra-low power silicon neuron
  circuit for extreme-edge neuromorphic intelligence,'' in \emph{International
  Conference on Electronics, Circuits, and Systems, ({ICECS}), 2019}, 11 2019,
  pp. 458--461.

\bibitem{Mead89}
C.~Mead, \emph{Analog {VLSI} and Neural Systems}.\hskip 1em plus 0.5em minus
  0.4em\relax Reading, MA: Addison-Wesley, 1989.

\bibitem{Vittoz96}
E.~Vittoz, ``{A}nalog {VLSI} implementation of neural networks,'' in
  \emph{{H}andbook of {N}eural {C}omputation}, E.~Fiesler and R.~Beale,
  Eds.\hskip 1em plus 0.5em minus 0.4em\relax Oxford and Bristol: {O}xford
  {U}niversity {P}ress and {I}nstitute of {P}hysics Publishing, 1996, ch. E1.3.

\bibitem{Liu_etal02a}
S.-C. Liu, J.~Kramer, G.~Indiveri, T.~Delbruck, and R.~Douglas, \emph{Analog
  {VLSI}:Circuits and Principles}.\hskip 1em plus 0.5em minus 0.4em\relax MIT
  Press, 2002.

\bibitem{Schemmel_etal17}
J.~Schemmel, L.~Kriener, P.~M{\"u}ller, and K.~Meier, ``An accelerated analog
  neuromorphic hardware system emulating {NMDA}- and calcium-based non-linear
  dendrites,'' in \emph{Neural Networks (IJCNN), 2017 International Joint
  Conference on}.\hskip 1em plus 0.5em minus 0.4em\relax IEEE, 2017, pp.
  2217--2226.

\bibitem{Mayr_etal16}
C.~Mayr, J.~Partzsch, M.~Noack, S.~H{\"a}nzsche, S.~Scholze, S.~H{\"o}ppner,
  G.~Ellguth, and R.~S. 2, ``A biological-realtime neuromorphic system in
  28\,nm {CMOS} using low-leakage switched capacitor circuits,'' \emph{{IEEE}
  Transactions on Biomedical circuits and systems}, vol.~10, no.~1, pp. 243 --
  254, 2016.

\bibitem{Folowosele_etal09}
F.~Folowosele, R.~Etienne-Cummings, and T.~Hamilton, ``A {CMOS} switched
  capacitor implementation of the {M}ihalas-{N}iebur neuron,'' in
  \emph{Biomedical Circuits and Systems Conference, ({BioCAS}), 2009}.\hskip
  1em plus 0.5em minus 0.4em\relax IEEE, Nov. 2009, pp. 105--108.

\bibitem{Folowosele_etal09a}
F.~Folowosele, A.~Harrison, A.~Cassidy, A.~Andreou, R.~Etienne-Cummings,
  S.~Mihalas, Niebur, and T.~Hamilton, ``A switched capacitor implementation of
  the generalized linear integrate-and-fire neuron,'' in \emph{International
  Symposium on Circuits and Systems, ({ISCAS}), 2009}.\hskip 1em plus 0.5em
  minus 0.4em\relax IEEE, May 2009, pp. 2149--2152.

\bibitem{Bartolozzi_etal06}
C.~Bartolozzi, S.~Mitra, and G.~Indiveri, ``An ultra low power current--mode
  filter for neuromorphic systems and biomedical signal processing,'' in
  \emph{Biomedical Circuits and Systems Conference, ({BioCAS}), 2006}.\hskip
  1em plus 0.5em minus 0.4em\relax IEEE, 2006, pp. 130--133.

\bibitem{Bartolozzi_Indiveri07a}
C.~Bartolozzi and G.~Indiveri, ``Synaptic dynamics in analog {VLSI},''
  \emph{Neural Computation}, vol.~19, no.~10, pp. 2581--2603, Oct 2007.

\bibitem{Saxena_Baker08}
V.~Saxena and R.~J. Baker, ``Compensation of {CMOS} op-amps using split-length
  transistors,'' in \emph{Circuits and Systems ({MWSCAS}), 2008 {IEEE} 51st
  International Midwest Symposium on}.\hskip 1em plus 0.5em minus 0.4em\relax
  IEEE, 2008, pp. 109--112.

\bibitem{Indiveri03a}
G.~Indiveri, ``A low-power adaptive integrate-and-fire neuron circuit,'' in
  \emph{International Symposium on Circuits and Systems, ({ISCAS}), 2003},
  vol.~IV.\hskip 1em plus 0.5em minus 0.4em\relax IEEE, May 2003, pp. 820--823.

\bibitem{Livi_Indiveri09}
P.~Livi and G.~Indiveri, ``A current-mode conductance-based silicon neuron for
  address-event neuromorphic systems,'' in \emph{International Symposium on
  Circuits and Systems, ({ISCAS}), 2009}.\hskip 1em plus 0.5em minus
  0.4em\relax IEEE, May 2009, pp. 2898--2901.

\bibitem{Indiveri_etal11}
\BIBentryALTinterwordspacing
G.~Indiveri, B.~Linares-Barranco, T.~Hamilton, A.~van Schaik,
  R.~Etienne-Cummings, T.~Delbruck, S.-C. Liu, P.~Dudek, P.~H{\"a}fliger,
  S.~Renaud, J.~Schemmel, G.~Cauwenberghs, J.~Arthur, K.~Hynna, F.~Folowosele,
  S.~Saighi, T.~Serrano-Gotarredona, J.~Wijekoon, Y.~Wang, and K.~Boahen,
  ``Neuromorphic silicon neuron circuits,'' \emph{Frontiers in Neuroscience},
  vol.~5, pp. 1--23, 2011. [Online]. Available:
  \url{http://www.frontiersin.org/Neuromorphic_Engineering/10.3389/fnins.2011.00073/abstract}
\BIBentrySTDinterwordspacing

\bibitem{Brette_Gerstner05}
R.~Brette and W.~Gerstner, ``Adaptive exponential integrate-and-fire model as
  an effective description of neuronal activity,'' \emph{Journal of
  neurophysiology}, vol.~94, no.~5, pp. 3637--3642, 2005.

\bibitem{Naud_etal08}
R.~Naud, N.~Marcille, C.~Clopath, and W.~Gerstner, ``Firing patterns in the
  adaptive exponential integrate-and-fire model,'' \emph{Biological
  Cybernetics}, vol.~99, no. 4--5, pp. 335--347, November 2008.

\bibitem{Izhikevich04}
E.~Izhikevich, ``Which model to use for cortical spiking neurons?''
  \emph{{IEEE} Transactions on Neural Networks}, vol.~15, no.~5, pp.
  1063--1070, September 2004.

\bibitem{Ha_Cheong17}
G.~E. Ha and E.~Cheong, ``Spike frequency adaptation in neurons of the central
  nervous system,'' \emph{Experimental neurobiology}, vol.~26, no.~4, pp.
  179--185, 2017.

\bibitem{Indiveri_etal10}
G.~Indiveri, F.~Stefanini, and E.~Chicca, ``Spike-based learning with a
  generalized integrate and fire silicon neuron,'' in \emph{International
  Symposium on Circuits and Systems, ({ISCAS}), 2010}.\hskip 1em plus 0.5em
  minus 0.4em\relax Paris, France: IEEE, 2010, pp. 1951--1954.

\bibitem{Qiao_etal17}
N.~Qiao, C.~Bartolozzi, and G.~Indiveri, ``An ultralow leakage synaptic scaling
  homeostatic plasticity circuit with configurable time scales up to 100 ks,''
  \emph{{IEEE} Transactions on Biomedical Circuits and Systems}, 2017.

\bibitem{Qiao_Indiveri16}
N.~Qiao and G.~Indiveri, ``Scaling mixed-signal neuromorphic processors to 28nm
  {FD-SOI} technologies,'' in \emph{Biomedical Circuits and Systems Conference,
  ({BioCAS}), 2016}.\hskip 1em plus 0.5em minus 0.4em\relax IEEE, 2016, pp.
  552--555.

\bibitem{Nair_Indiveri19}
M.~V. {Nair} and G.~{Indiveri}, ``An ultra-low power sigma-delta neuron
  circuit,'' in \emph{2019 IEEE International Symposium on Circuits and Systems
  (ISCAS)}, May 2019, pp. 1--5.

\end{thebibliography}

\begin{IEEEbiography}[{\includegraphics[width=1in,height=1.25in,clip,keepaspectratio]{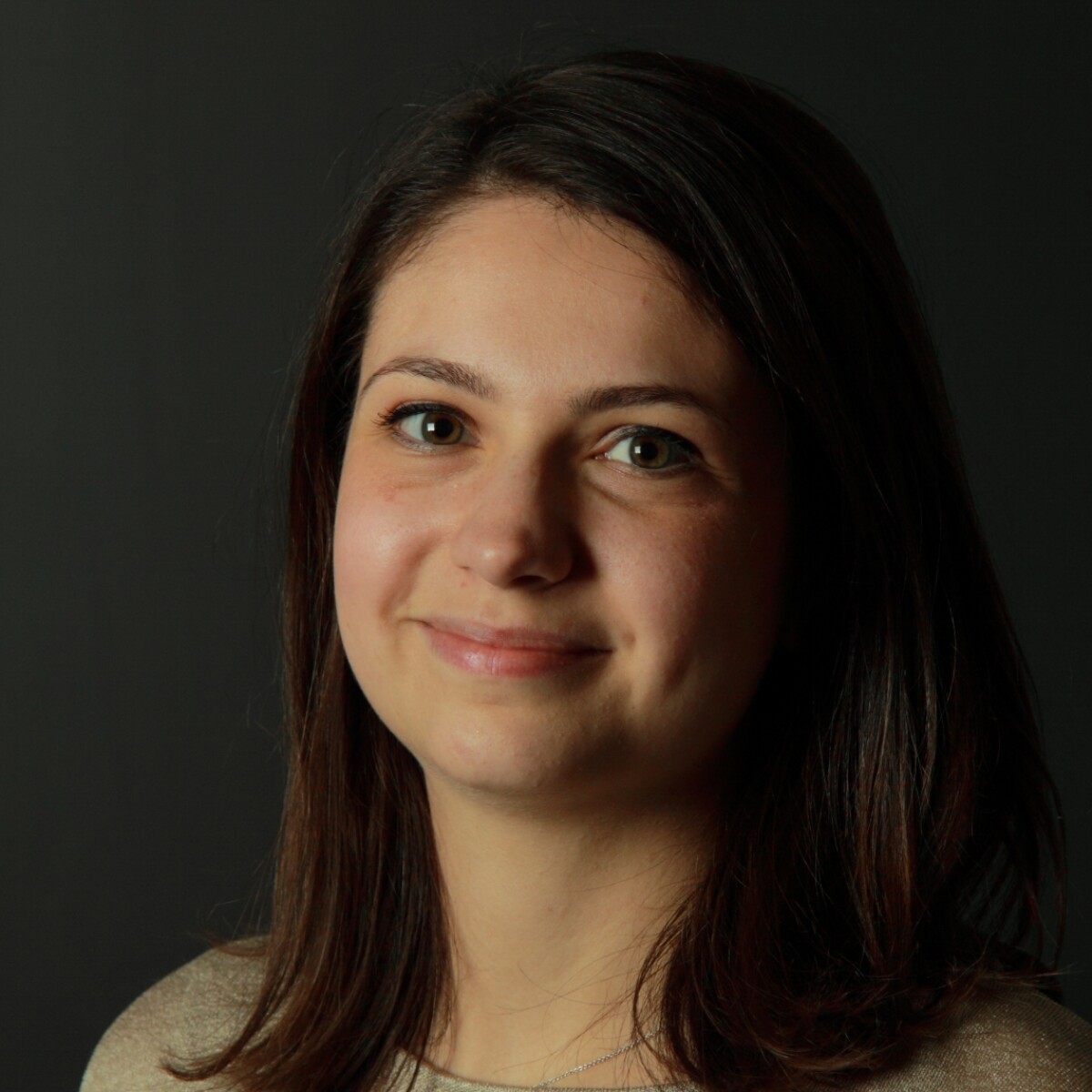}}]{Arianna Rubino}
received her B.Sc degree in 2017 from Politecnico di Milano, Italy in biomedical engineering and her M.S degree in 2019 in biomedical engineering with specialization in bioelectronics from the Swiss Federal Institute of Technology in Zurich, Switzerland. Since September 2019 she is working toward the Ph.D. degree at the Institute of Neuroinformatics, University of
Zurich and ETH Zurich, Zurich,
Switzerland. Her research interests include the design of ultra-low power mixed-signal circuits for neuromorphic edge computing and biomedical applications using advanced transistor processes and the implementation of biologically plausible learning algorithms on-chip.
\end{IEEEbiography}

\begin{IEEEbiography}[{\includegraphics[width=1in,height=1.25in,clip,keepaspectratio]{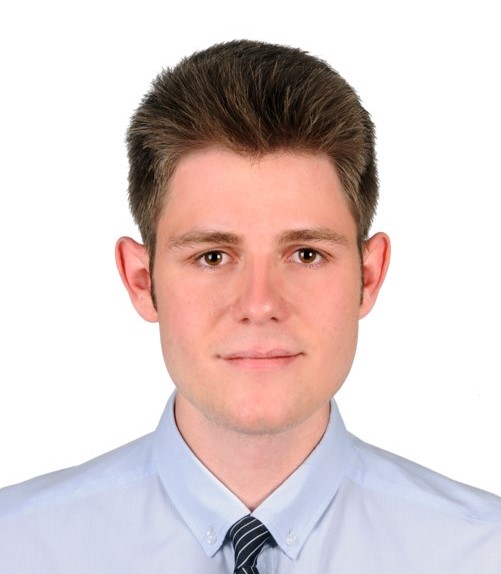}}]{Can Livanelioglu}
received his B.Sc degree in Electrical and Electronics Engineering from Middle East Technical University, Ankara, Turkey in 2019, with specialization in Electronics and Biomedical. He is now pursuing his M.Sc. degree in Biomedical Engineering with specialization in bioelectronics at ETH Zurich, Switzerland and working at the Institute of Neuroinformatics, ETH Zurich and University of Zurich, Switzerland. His research interests include the design of analog/digital integrated circuits, solid state electronics, novel semiconductor device architectures and neuromorphic structures.
\end{IEEEbiography}

\begin{IEEEbiography}[{\includegraphics[width=1in,height=1.25in,clip,keepaspectratio]{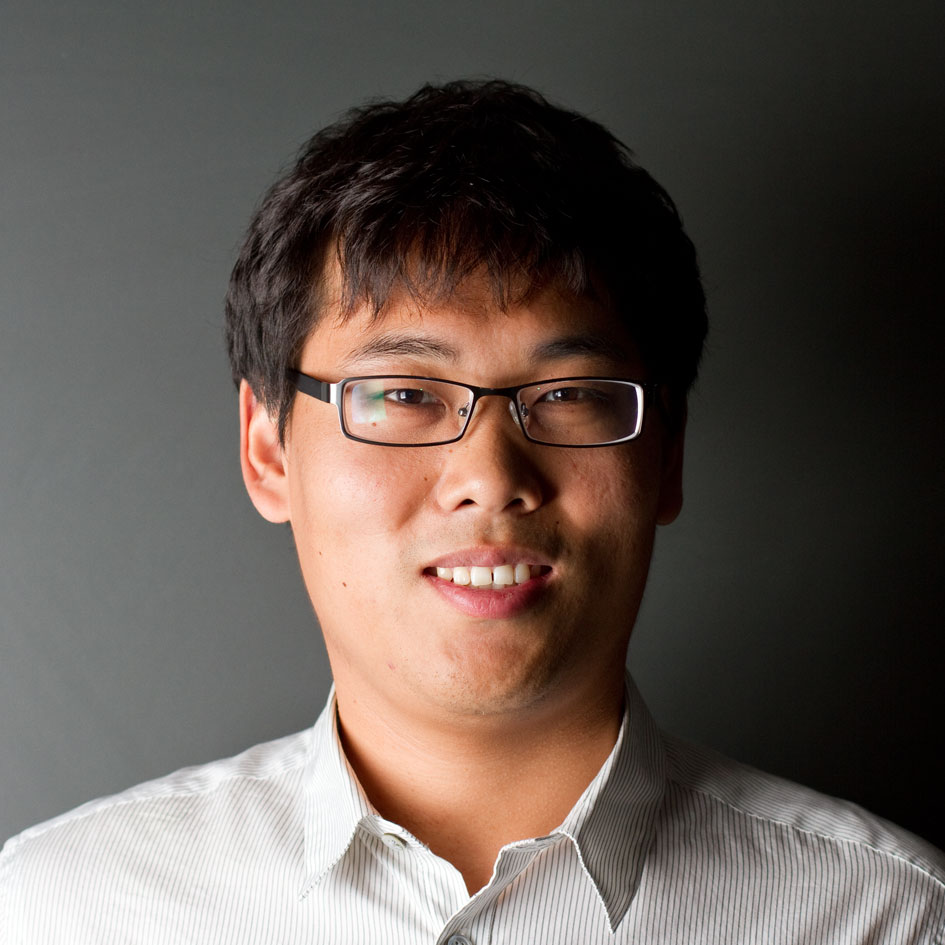}}]{Ning Qiao}
received the bachelor’s degree in microelectronics and solid-state electronics from Xi’an Jiaotong University, Xi’an, China, in 2006, and the Ph.D. degree in microelectronics from the Institute of Semiconductors, Chinese Academy of Sciences, China, in 2012, with a focus on ultra-low-power low-noise mixed-signal circuits in SOI process. He joined the Institute of Neuroinformatics, University of Zurich and ETH Zürich, Switzerland, as a Post-Doctoral Researcher, in 2012, where he is involved in developing mixed-signal multicore neuromorphic VLSI circuits and systems. His current research interests include ultra-low-power subthreshold mixed-signal neuromorphic VLSI circuits and systems, parallel neuromorphic computing architectures, and fully asynchronous event-driven computing and communication circuits and systems.
\end{IEEEbiography}

\begin{IEEEbiography}[{\includegraphics[width=1in,height=1.25in,clip,keepaspectratio]{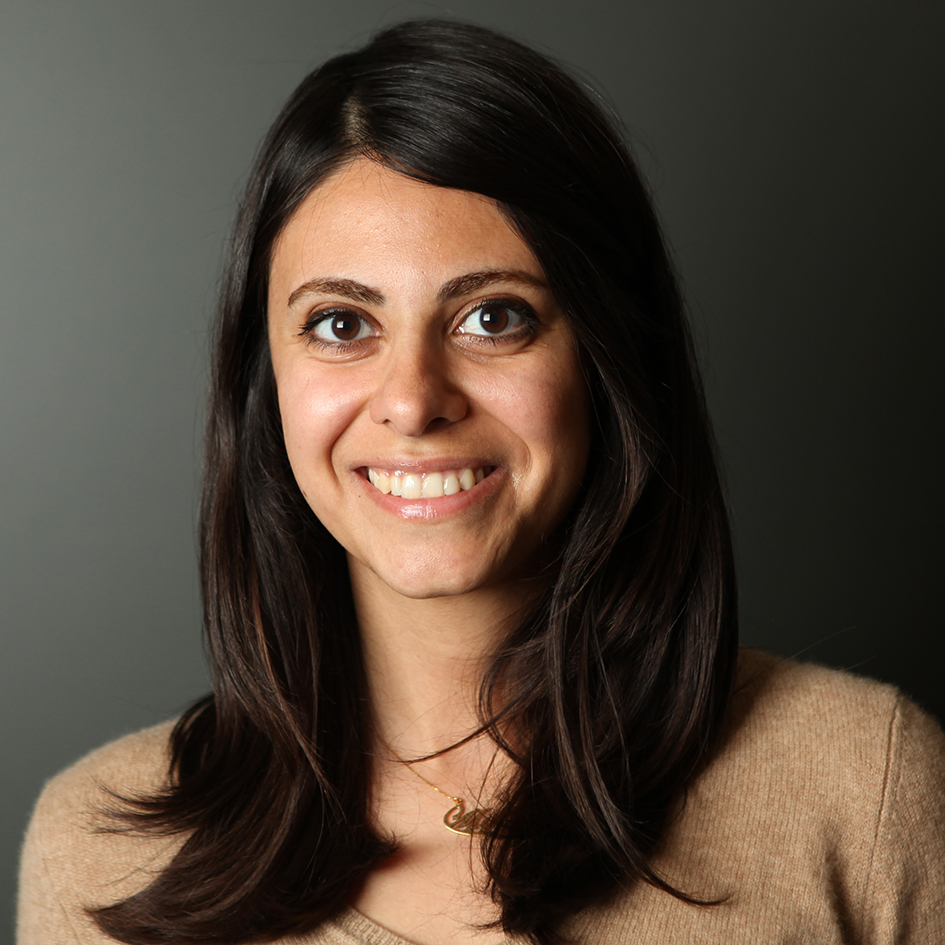}}]{Melika Payvand}
received her B.Sc degree in 2010 from University of Tehran, Iran in electrical engineering and her M.S. and Ph.D. degree in electrical and computer engineering from University of California Santa Barbara in 2012 and 2016 respectively. 
Currently, she is a post-doctorate researcher at the Institute of Neuroinformatics, University of Zurich and ETH Zurich. Her research activities and interest is in exploiting the physics of the computational substrate for real-time sensory processing. Specifically, she is interested in exploiting the physics of the computational substrate in event-based neuromorphic chips to enable low power and highly dense solutions for wearable and IoT applications.
\end{IEEEbiography}

\begin{IEEEbiography}[{\includegraphics[width=1in,height=1.25in,clip,keepaspectratio]{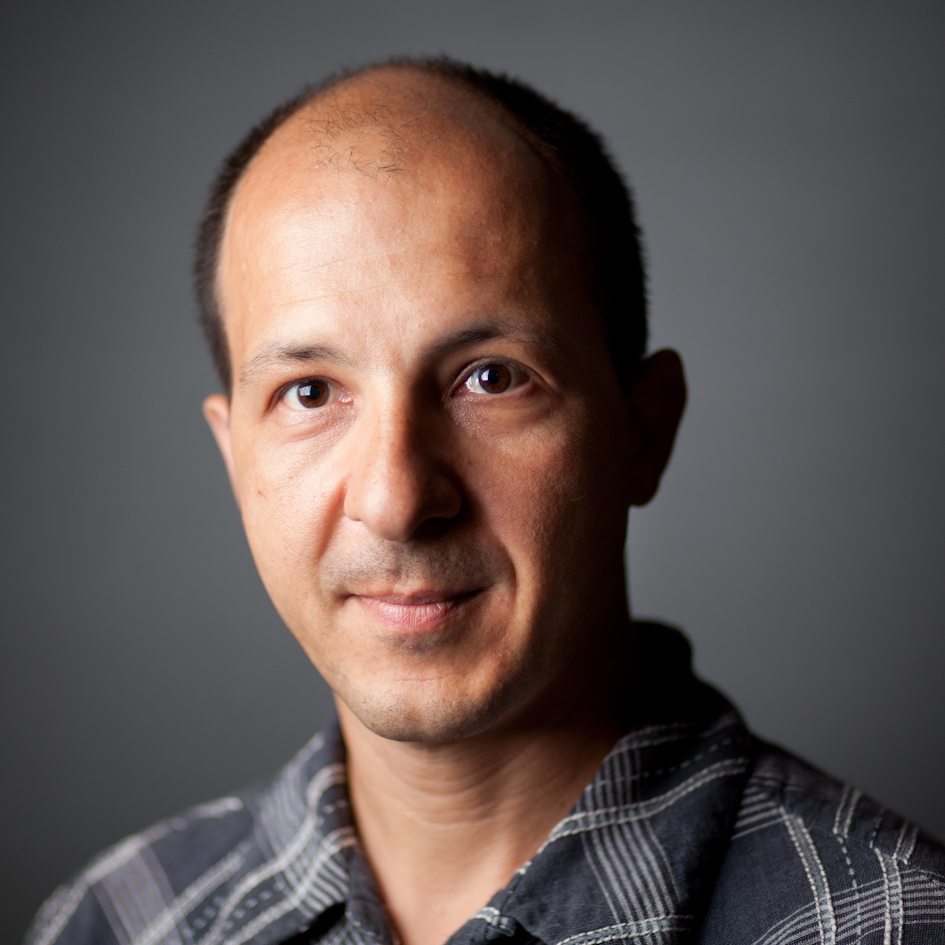}}]{Giacomo Indiveri}
is a dual Professor at the Faculty of Science of the University of Zurich and at Department of Information Technology and Electrical Engineering of ETH Zurich, Switzerland. He is the director of the Institute of Neuroinformatics (INI) of the University of Zurich and ETH Zurich. He obtained an M.Sc. degree in electrical engineering in 1992 and a Ph.D. degree in computer science from the University of Genoa, Italy in 2004. He was a post-doctoral research fellow in the Division of Biology at Caltech and at the Institute of Neuroinformatics of the University of Zurich and ETH Zurich. He was awarded an ERC Starting Grant on "Neuromorphic processors" in 2011 and an ERC Consolidator Grant on neuromorphic cognitive agents in 2016. His research interests lie in the study of neural computation, with a particular focus on spike-based learning and selective attention mechanisms. His research and development activities focus on the full custom hardware implementation of real-time sensory-motor systems using analog/digital neuromorphic circuits and emerging memory technologies. Dr. Indiveri is a member of several technical committees of the IEEE Circuits and Systems Society and a Fellow of the European Research Council.
\end{IEEEbiography}

\end{document}